\documentclass[showpacs,twocolumn,aps,floatfix,superscriptaddress]{revtex4-1}
\usepackage{amsmath,amssymb,eucal,graphicx,bm}

\begin{document}
\title{A Light Impurity in an Equilibrium Gas}
\author{L. D'Alessio}
\affiliation{Department of Physics, Boston University, Boston, MA 02215, USA}
\author{P. L. Krapivsky}
\affiliation{Department of Physics, Boston University, Boston, MA 02215, USA}
\affiliation{Institut de Physique Th\'eorique CEA, IPhT, F-91191 Gif-sur-Yvette, France}
\begin{abstract}
We investigate the evolution of a light impurity particle in a Lorentz gas where the background atoms are in thermal equilibrium. As in the standard Lorentz gas, we assume that the particle is negligibly light in comparison with the background atoms.  The thermal motion of atoms causes the average particle speed  to grow. In the case of the hard-sphere particle-atom interaction, the temporal growth is ballistic, while generally it is sub-linear. For the particle-atom potential that diverges as $r^{-\lambda}$ in the small separation limit, the average particle speed grows as $t^{\lambda/(2(d-1)+\lambda)}$ in $d$ dimensions. 
The particle displacement exhibits a universal growth, linear in time and the average (thermal) speed of the atoms. Surprisingly, the asymptotic growth is independent of the gas density  and the particle-atom interaction. The velocity and position distributions approach universal scaling forms which are non-Gaussian. We determine the velocity distribution in arbitrary dimension and for arbitrary interaction exponent $\lambda$.  For the hard-sphere particle-atom interaction, we compute the position distribution and the joint velocity-position distribution. 
\end{abstract}
\maketitle

\section{Introduction} 

The goal of this work is to investigate the behavior of an impurity particle (particle in short) in a monoatomic gas. We focus on the limit when the particle is negligibly light in comparison with background atoms. In other words, the particle is affected by collisions with atoms, while atoms do not ``feel'' the presence of the particle. We want to understand the evolution of the particle velocity and displacement distribution. 

The problem is a natural generalization of the standard Lorentz gas 
\cite{LG,H74,fluid,dorfman,LG-math,book} where scatters are assumed to be immobile. 
The speed of the particle remains constant in the framework of the Lorentz model. 
In our model the behavior is completely different and can be simply understood using arguments from the equipartition theorem (when the background gas has a positive temperature the average speed of the particle increases without a bound since the particle ``tries'' to reach an equilibrium with the background atoms). 

The problem is also reminiscent of the model originally proposed by Fermi \cite{F}, and later refined by Ulam \cite{U}, to explain the acceleration of interstellar particles and cosmic rays. Fermi's acceleration mechanism has been mostly studied using methods of dynamical systems (see \cite{LL} and references therein); an application of kinetic theory to Fermi's mechanism has been presented in \cite{italy}. 

Here we analyze the behavior of the light particle in an equilibrium gas using the Boltzmann equation framework. The Boltzmann equation \cite{BOLT} is the basic tool in elucidating the properties of transport phenomena. The non-linear integro-differential Boltzmann equation is so formidable, however, that apart from the equilibrium Maxwell-Boltzmann distribution \cite{MAX} there are essentially no solutions to the Boltzmann equation \cite{TM80}. The standard Lorentz gas model where a point particle is elastically scattered by immobile hard spheres is described by the Lorentz-Boltzmann equation \cite{LG} which is linear and, not surprisingly, amenable to analytical treatments. The Lorentz gas has played an outstanding role in concrete calculations (e.g. of the diffusion coefficient) and in the conceptual development of kinetic theory \cite{H74,fluid}. Yet the very applicability of the Boltzmann framework to the Lorentz gas is questionable --- when the scatters are fixed, the molecular chaos assumption underlying the Boltzmann equation cannot be justified \cite{H74,fluid,dorfman,LG-math,book}. 

If, however, the background atoms move and collide with each other, the molecular chaos assumption holds in the dilute limit and the (properly generalized) Lorentz-Boltzmann equation must be applicable as long as the mass of the particle is infinitesimally small so that it does not affect the motion of atoms.  Moreover, since the (average) particle speed continues to grow, it eventually greatly exceeds the typical velocities of background atoms. This allows to simplify the most difficult term in the Boltzmann equation, the so-called collision integral; mathematically, an integral operator becomes a differential one and the integro-differential Lorentz-Boltzmann equation reduces to a partial differential equation. 

The unlimited velocity growth suggests that the particle velocity distribution approaches a scaling form. The scaled velocity distribution satisfies an ordinary differential equation (Sects.~\ref{1D}--\ref{mono}) which admits a simple solution; for the hard-sphere atoms, the scaled velocity distribution is exponential (Sects.~\ref{1D}--\ref{HSG}).  The Boltzmann equation approach also describes the spatial distribution of the particle, yet extracting the density distribution is much more difficult as it does not obey a closed equation, so one must rely on the joint distribution function that simultaneously describes the probability density for the position and velocity. In Sec.~\ref{displacement} we outline the evolution of the displacement using heuristic arguments and exact calculations in one dimension based on the velocity correlation functions. In Sect.~\ref{VPD} we derive kinetic equations describing the joint distribution in the long-time limit. In Sect.~\ref{moment_approach} we investigate the density profile of the hard-sphere gas by utilizing the moment approach and in Sect.~\ref{full_solution} we compute the joint distribution. We report the results of numerical simulations in Sec.~\ref{numerics} and summarize our findings in Sect.~\ref{summ}.
 
\section{One Dimension}
\label{1D}

As a warm-up, consider the one-dimensional case. This may appear physically dubious as the particle is caged between two adjacent atoms, so the molecular chaos assumption (that is, the lack of correlations between pre-collision velocities) underlying the Boltzmann approach is certainly invalid in one dimension. A Boltzmann equation, however, makes sense if we consider the situation when in each collision the scattering occurs with a certain probability (otherwise the particle and an atom just pass through each other). This one-dimensional Boltzmann equation sheds light on the three-dimensional case. Therefore it has been proven useful as a toy model and it has been studied in a number of one-dimensional settings (see e.g. \cite{R,GP,PS,AP,book}). 

The Boltzmann equation for the particle velocity distribution $f(v,t)$ reads 
\begin{equation}
\label{B-1d}
\frac{\partial f(v,t)}{\partial t}=\int_{-\infty}^\infty du\,|v-u|\,P(u)[f(2u-v,t)-f(v,t)]
\end{equation}
Here
\begin{equation}
\label{MB-1d}
P(u) = \frac{\rho}{\sqrt{2\pi T}}\,e^{-u^2/2T}
\end{equation}
is the equilibrium velocity distribution of the background atoms corresponding to temperature $T$ (we set the atomic mass to unity). We shall see, however, that we do not need the detailed form \eqref{MB-1d} of the equilibrium Maxwell-Boltzmann distribution. To establish the asymptotic behavior of $f(v,t)$ it is sufficient to assume that $P(u)$ is an even function, $P(u)=P(-u)$. Even a weaker condition that the average velocity of atoms vanishes,
\begin{equation}
\label{Pu-1}
\int_{-\infty}^\infty du\,u\,P(u)=0,
\end{equation}
suffices. Whenever \eqref{Pu-1} holds, the long-time behavior depends only on the second moment of $P(u)$ which essentially defines the temperature: 
\begin{equation}
\label{Pu-2}
\int_{-\infty}^\infty du\,u^2\,P(u)=\rho T
\end{equation}
We shall see that in the long-time, more precisely when
\begin{equation}
\label{long-time}
t\gg \rho^{-1}T^{-1/2}
\end{equation}
the Boltzmann equation \eqref{B-1d} for the particle velocity distribution simplifies to
\begin{equation}
\label{B-1d-scaling}
\frac{\partial f}{\partial \tau}=\frac{\partial f}{\partial v}+v\,\frac{\partial^2 f}{\partial v^2}\,,\quad 
\tau=2\rho T t
\end{equation}
This kinetic equation admits the scaling solution
\begin{equation}
\label{scaling-1d}
f(v,t) = \frac{1}{2\tau}\,e^{-|v|/\tau}
\end{equation}

To derive \eqref{B-1d-scaling}--\eqref{scaling-1d} we first simplify the collision integral in Eq.~\eqref{B-1d} in the $t\to\infty$ limit. Since  $f(v,t)=f(-v,t)$, it suffices to investigate the $v>0$ region \cite{even}. Moreover we can replace $|v-u|$ by $v-u$ since the region $v<u$ where the replacement is invalid provides a negligible contribution in the long-time limit: $P(u)$ is very small in this region. More precisely, the above simplification applies if the average speed of atoms $\langle u\rangle \sim \sqrt{T}$ is much smaller than the particle velocity $v$. This is our working assumption which will be checked a posteriori. When $\langle u\rangle\ll v$ we can additionally expand $f(2u-v)$ that appears in the collision integral in Eq.~\eqref{B-1d} into a Taylor series
\begin{eqnarray*}
f(2u-v) &=& 
f(v)-2u\,\frac{\partial f(v)}{\partial v}+2u^2\,\frac{\partial^2 f(v)}{\partial v^2}\\
&-&\frac{(2u)^3}{3!}\,\frac{\partial^3 f(v)}{\partial v^3}+
\frac{(2u)^4}{4!}\,\frac{\partial^4 f(v)}{\partial v^4}+
\ldots
\end{eqnarray*}
Plugging this expansion into Eq.~\eqref{B-1d} and computing the integrals over $u$ we obtain
\begin{equation}
\label{1d-long}
\frac{\partial f}{\partial \tau}=
\frac{\partial f}{\partial v}+v\,\frac{\partial^2 f}{\partial v^2}
+2T\left(\frac{\partial^3 f}{\partial v^3}+v\,\frac{\partial^4 f}{\partial v^4}\right)+\ldots
\end{equation}
In computing the integrals leading to the first two terms on the right-hand side of \eqref{1d-long} it suffices to use the integral relations \eqref{Pu-1}--\eqref{Pu-2}. The next two terms are obtained using the integral relations
\begin{equation}
\label{Pu-34}
\int_{-\infty}^\infty du\,u^3\,P(u)=0, \quad 
\int_{-\infty}^\infty du\,u^4\,P(u)=3\rho T^2
\end{equation}
The first relation in \eqref{Pu-34} is valid for any symmetric velocity distribution, $P(u)=P(-u)$, while the second is derived from the equilibrium Maxwell-Boltzmann distribution \eqref{MB-1d}. 

The first two terms on the right-hand side of \eqref{1d-long} scale as $\tau^{-1}$, the next two terms scale as $T\tau^{-3}$, so they are asymptotically negligible when $\tau\gg\sqrt{T}$, that is, the average particle speed greatly exceeds the average speed of atoms. The two following terms [which haven't been displayed in \eqref{1d-long}] contains $T^2\frac{\partial^5 f}{\partial v^5}$ and 
$T^2v\frac{\partial^6 f}{\partial v^6}$, so they scale as $T^2\tau^{-5}$ and therefore they are even smaller. Thus in the $\tau\gg\sqrt{T}$ limit (which is given by Eq.~\eqref{long-time} in the original variables), Eq.~\eqref{1d-long} indeed reduces to Eq.~\eqref{B-1d-scaling} in the leading order. 

The form of equation \eqref{B-1d-scaling} suggests to seek the scaling solution of the form 
\begin{equation}
\label{scaling-1d-form}
f(v,\tau)= \tau^{-1}\Phi(w), \quad w = v/\tau
\end{equation}
Plugging \eqref{scaling-1d-form} into \eqref{B-1d-scaling} we obtain an ordinary differential equation for $\Phi(w)$ which is solved to yield $\Phi(w)=C\,e^{-w}$. Recalling that the particle velocity distribution is even and using the normalization condition $\int dv\,f(v,t)=1$ fixes the amplitude $C=1/2$ and leads to the announced result \eqref{scaling-1d}.

Having determined the scaling solution \eqref{scaling-1d}, we would like to understand if any arbitrary function $f(v, t)$ approach the scaling solution  \eqref{scaling-1d} in the long time limit. The answer to this question is presumably affirmative, at least when the initial velocity distribution $f(v,t=0)$ quickly decays when $|v|\to\infty$. Yet to prove this assertion even for simplest initial velocity distributions like $f(v,t=0)=\delta(v)$ is hard. Analytical arguments showing that the scaling solution \eqref{scaling-1d} is indeed an attractor are presented in Appendix~\ref{approach}. 

\section{Hard-sphere gas}
\label{HSG}

Consider now the most natural three-dimensional situation and assume that atoms are hard spheres of radius $a$. We ignore both the mass and the size of the particle. The latter assumption is not crucial --- if the particle is a sphere of radius $b$, it suffices to replace $a$ by $a+b$ in the following formulae. 

We again employ the Boltzmann equation approach. This framework is applicable only in the diluted limit; for the hard-sphere gas, this means that the volume fraction occupied by atoms is small: $\rho\times \frac{4\pi}{3}\,a^3\ll 1$ (here $\rho$ is the number density of background atoms). 

The Boltzmann equation reads
\begin{equation}
\label{B-3d}
\frac{\partial f({\bf v},t)}{\partial t}=\int d{\bf u}\,P({\bf u})\,ga^2\!
\int\! \mathcal{D}{\bf e}\,[f({\bf v}',t)-f({\bf v},t)]
\end{equation}
Here ${\bf e}$ is the unit vector pointing to the position of the particle at the moment when it hits the sphere. The post-collision velocity ${\bf v}'$  of the particle can be expressed via ${\bf v}, {\bf e}$, and the relative velocity ${\bf g}={\bf u}-{\bf v}$:
\begin{equation}
\label{vv}
{\bf v}' = {\bf v} + 2{\bf e}({\bf g}\cdot{\bf e})
\end{equation}
In Eq.~\eqref{B-3d} we have also used the shorthand notation $\mathcal{D}{\bf e}$ for the integration measure over angular coordinates. For the hard-sphere gas, this integration measure reads \cite{fluid}
\begin{equation}
\label{De}
\mathcal{D}{\bf e} = \frac{({\bf g}\cdot{\bf e})}{g}\,
\theta({\bf g}\cdot{\bf e})\,d^2{\bf e}
\end{equation}
In the above expression $\theta(\cdot)$ is the Heaviside step function and $d^2{\bf e}$ is the standard angular integration measure. 

To simplify the Boltzmann equation \eqref{B-3d} we shall proceed as in one dimension. Since the particle velocity distribution is (asymptotically) isotropic, let us treat $f({\bf v})$ as a function of $V=v^2=({\bf v}\cdot {\bf v})$. Squaring \eqref{vv} we get
\begin{equation*}
V' = V + 4({\bf v}\cdot{\bf e})({\bf g}\cdot{\bf e})+4({\bf g}\cdot{\bf e})^2
=V+4({\bf u}\cdot{\bf e})({\bf g}\cdot{\bf e})
\end{equation*}
Using this result and expanding $f({\bf v}')=f(V')$ into a Taylor series we obtain
\begin{equation*}
f(V')=f(V)+4({\bf u}\cdot{\bf e})({\bf g}\cdot{\bf e})\,\frac{\partial f}{\partial V}
+ 8({\bf u}\cdot{\bf e})^2({\bf g}\cdot{\bf e})^2\,\frac{\partial^2 f}{\partial V^2}+\ldots
\end{equation*}
Using this expansion we simplify \eqref{B-3d} to
\begin{equation}
\label{long-eq}
\begin{split}
\frac{\partial f}{\partial t} &= 4\,\frac{\partial f}{\partial V}\! \int\! d{\bf u}\,P({\bf u})\, 
ga^2\!\!\int \mathcal{D}{\bf e}\,({\bf u}\cdot{\bf e})({\bf g}\cdot{\bf e})\\
&+ 8\,\frac{\partial^2 f}{\partial V^2}\! \int\! d{\bf u}\,P({\bf u})\,ga^2\!\!
\int \mathcal{D}{\bf e}\,({\bf u}\cdot{\bf e})^2({\bf g}\cdot{\bf e})^2
\end{split}
\end{equation}
As in the one-dimensional case, it suffices to keep only the terms with the first and second order derivatives in $V$; the terms with higher order derivatives are asymptotically negligible. The angular integrals in Eq.~\eqref{long-eq} are computed [see Appendix~\ref{app}] to yield 
\begin{subequations}
\label{integrals}
\begin{align}
\int \mathcal{D}{\bf e}\,({\bf u}\cdot{\bf e})({\bf g}\cdot{\bf e})&= 
\frac{\pi}{2}\, ({\bf g}\cdot{\bf u})
\label{integralA}\\
\int \mathcal{D}{\bf e}\,({\bf u}\cdot{\bf e})^2({\bf g}\cdot{\bf e})^2
&= \frac{\pi}{12}\, \left[3 ({\bf g}\cdot{\bf u})^2 +g^2 u^2\right]
\label{integralB}
\end{align}
\end{subequations}
Inserting \eqref{integralA}--\eqref{integralB} into Eq.~\eqref{long-eq} we obtain
\begin{equation}
\label{long2}
\begin{split}
\frac{1}{2\pi a^2}\,
\frac{\partial f}{\partial t} &=
\frac{v}{3}\,\frac{\partial^2 f}{\partial V^2} \int d{\bf u}\,P({\bf u})
\left[3 ({\bf v}\cdot{\bf u})^2 +v^2 u^2\right]\\
&\quad+ \frac{\partial f}{\partial V}\! \int\! d{\bf u}\,P({\bf u})
\,g ({\bf g}\cdot{\bf u})
\end{split}
\end{equation}
In the first integral we already replaced ${\bf g}$ by $-{\bf v}$ which is correct in the leading order. In the second integral we should be more careful. We write
\begin{equation*}
g ({\bf g}\cdot{\bf u}) = -v ({\bf v}\cdot{\bf u}) 
+ v^{-1}\left[({\bf v}\cdot{\bf u})^2 + v^2u^2\right]+\ldots
\end{equation*}
The integral that contains the leading term vanishes since 
$\int d{\bf u}\,P({\bf u})\,{\bf u}=0$. Thus Eq.~\eqref{long2} becomes
\begin{equation}
\label{long3}
\begin{split}
\frac{1}{2\pi a^2}\,
\frac{\partial f}{\partial t} &=
\frac{v}{3}\,\frac{\partial^2 f}{\partial V^2} \int d{\bf u}\,P({\bf u})
\left[3 ({\bf v}\cdot{\bf u})^2 +v^2 u^2\right]\\
&\quad + \frac{1}{v}\,\frac{\partial f}{\partial V} 
\int d{\bf u}\,P({\bf u})\left[({\bf v}\cdot{\bf u})^2 +v^2 u^2\right]
\end{split}
\end{equation}
Using relations
\begin{equation}
\label{T-3d}
\int d{\bf u}\,P({\bf u})\,u^2 = 3\rho T, \quad 
\int d{\bf u}\,P({\bf u})\,({\bf v}\cdot{\bf u})^2 = v^2\rho T
\end{equation}
we recast \eqref{long3} into 
\begin{equation}
\label{long4}
\frac{\partial f}{\partial \tau} = 8v\,\frac{\partial f}{\partial V} 
+ 4v^3\,\frac{\partial^2 f}{\partial V^2}\,,\quad \tau = \pi a^2\rho T t
\end{equation}
Since $V=v^2$, we have
\begin{equation}
\label{VV}
\frac{\partial }{\partial V} = \frac{1}{2v}\,\frac{\partial }{\partial v}\,,\quad 
\frac{\partial^2}{\partial V^2} = -\frac{1}{4v^3}\,\frac{\partial}{\partial v}
+ \frac{1}{4v^2}\,\frac{\partial^2}{\partial v^2}
\end{equation}
Using these identities we re-write \eqref{long4} as
\begin{equation}
\label{long5}
\frac{\partial f}{\partial \tau} = 3\,\frac{\partial f}{\partial v} 
+ v\,\frac{\partial^2 f}{\partial v^2}
\end{equation}
This kinetic equation admits the scaling solution 
\begin{equation}
\label{scaling-3d}
f(v,t) = \frac{1}{8\pi \tau^3}\,e^{-v/\tau}
\end{equation}
For instance, the average speed of the particle is 
\begin{equation*}
\langle v\rangle = \int_0^\infty v\,f(v,t)\, 4\pi v^2\,dv = 3\tau
\end{equation*}
and more generally 
\begin{equation}
\label{vn-av}
\langle v^n\rangle = \frac{(n+2)!}{2}\,\tau^n
\end{equation}

The above analysis can be straightforwardly extended from three to $d$ dimensions. The results up to \eqref{long-eq} require obvious amendments, e.g. in equation \eqref{long-eq} we must replace $a^2$ by $a^{d-1}$. The integrals \eqref{integralA}--\eqref{integralB} become (see Appendix \ref{app})
\begin{subequations}
\begin{align}
\int\mathcal{D}{\bf e}\,(\mathbf{u}\cdot\mathbf{e})(\mathbf{g}\cdot\mathbf{e}) & =A(\mathbf{u}\cdot\mathbf{g})
\label{int_A}\\
\int\mathcal{D}{\bf e}\,(\mathbf{u}\cdot\mathbf{e})^{2}(\mathbf{g}\cdot\mathbf{e})^{2} & =\frac{dB-A}{d-1}\,(\mathbf{u}\cdot\mathbf{g})^{2}+\frac{A-B}{d-1}\, g^{2}u^{2}
\label{int_B}
\end{align}
\end{subequations}
where $A$, $B$ are constants defined by integrals: 
\begin{equation}
\label{AB:def}
A=\frac{1}{g^{2}}\int\mathcal{D}{\bf e}\,(\mathbf{g}\cdot\mathbf{e})^{2}\,,\quad
B=\frac{1}{g^{4}}\int\mathcal{D}{\bf e}\,(\mathbf{g}\cdot\mathbf{e})^{4}\,.
\end{equation}
The governing kinetic equation that generalizes Eq.~\eqref{long5} reads 
\begin{equation} 
\frac{\partial f}{\partial\tau}=d\,\frac{\partial f}{\partial v}+v\,\frac{\partial^{2}f}{\partial v^{2}}\,,\,\,\tau=2a^{d-1}A\rho Tt.
\label{final_v}
\end{equation}
Interestingly, in all dimensions the constant $B$ drops from the final equation; the constant $A$ is essentially irrelevant as it is absorbed into the new time variable $\tau$. 

Equation \eqref{final_v} is much simpler than Eq.~\eqref{B-3d} and it can be solved by employing the Laplace transform (see Appendix~\ref{exact}). 
The asymptotic solution of Eq.~\eqref{final_v} is again a pure exponential 
\begin{equation}
f = \left[\Omega_d\,\Gamma(d)\right]^{-1}\tau^{-d}\, e^{-v/\tau}
\label{scaling-vd}
\end{equation}
where $\Omega_d=\frac{2\pi^{d/2}}{\Gamma(d/2)}$ is the area of the unit sphere in $d$ dimensions. The constant in \eqref{scaling-vd} has been chosen to ensure the normalization: $\int d \mathbf{v} f({\bf v},t)=1$.

In two dimensions, Eqs.~\eqref{final_v}--\eqref{scaling-vd} have been derived in Ref.~\cite{italy} in the realm of a stochastic model for Fermi's acceleration. Even earlier, the exponential velocity distribution was found to occur in another stochastic model for Fermi's acceleration \cite{poland} in which a particle is bouncing in a container of fixed volume with boundaries deforming in a chaotic manner. In this case, the velocity distribution becomes exponential independently of the container's shape and the deformation protocol.

\section{Monoatomic gas}
\label{mono}

Consider now a general case of a monoatomic gas. It is then natural to assume that the interaction between the particle and an atom separated by distance $r$ can be described by a potential function $U(r)$. In the long time limit when the particle velocity becomes large, only the small $r$ behavior of the potential $U(r)$ matters. In this limit, the repulsion part of the interaction dominates and it usually diverges algebraically in the small separation limit
\begin{equation}
\label{Ur}
U(r)\simeq \epsilon \left(\frac{r_0}{r}\right)^\lambda
\end{equation}
as $r\to 0$. For example, $\lambda=12$ for the Lennard-Jones potential (in three dimensions).

To estimate interaction size $r_*$ we can use the criterion $U(r_*)\sim g^2$, from which we find $r_*$ and the cross section area $\sigma\sim r_*^{d-1}$:
\begin{equation*}
r_*\sim r_0  \left(\frac{\epsilon}{g^2}\right)^{1/\lambda}\,, \quad
\sigma_* \sim (r_0)^{d-1} \left(\frac{\epsilon}{g^2}\right)^{(d-1)/\lambda}
\end{equation*}
The term $ga^{d-1}\mathcal{D}{\bf e}$ characterizing the hard-sphere gas should be replaced by the term $g\sigma_* \mathcal{D}{\bf e}$ in the general case. In one dimension, the interaction law is irrelevant and the problem reduces to the hard-core interaction. In higher dimensions, the Boltzmann equation depends on the interaction exponent $\lambda$ as it contains the factor $g\sigma_* \sim g^{1-\gamma}$ with $\gamma=2(d-1)/\lambda$. In the long-time limit, the particle is very fast, so it is scattered only when it greatly approaches the atom, that is the separation is small and therefore the above analysis is asymptotically exact. Thus we must merely replace $g$ by $g^{1-\gamma}$ in the Lorentz-Boltzmann equation. This gives 
\begin{equation}
\label{B-3d-gen}
\frac{\partial f({\bf v})}{\partial t}=
\!\int\! d{\bf u}\,P({\bf u})\,g^{1-\gamma}\!
\int\! \mathcal{D}{\bf e}\,[f({\bf v}')-f({\bf v})]
\end{equation}
where we absorbed the $(r_0\epsilon^{1/\lambda})^{d-1}$ factor into the time variable. 

To simplify the Boltzmann equation \eqref{B-3d-gen} we repeat the same steps as for the hard-sphere gas to yield 
\begin{equation}
\label{long-mu}
\begin{split}
\frac{\partial f}{\partial t} & = 4\,\frac{\partial f}{\partial V}\! \int\! d{\bf u}\,P({\bf u})\, 
g^{1-\gamma}\!\!\int \mathcal{D}{\bf e}\,({\bf u}\cdot{\bf e})({\bf g}\cdot{\bf e})\\
& +8\,\frac{\partial^2 f}{\partial V^2}\! \int\! d{\bf u}\,P({\bf u})\,g^{1-\gamma}\!\!
\int \mathcal{D}{\bf e}\,({\bf u}\cdot{\bf e})^2({\bf g}\cdot{\bf e})^2
\end{split}
\end{equation}
where we have kept the terms with the first and second order derivatives in $V$ as asymptotically they provide the leading contribution. Computing the angular integrals [as in Section \ref{HSG} and Appendix~\ref{app}] we arrive at 
\begin{equation}
\label{long-mu-2}
\begin{split}
\frac{1}{4A}\,\frac{\partial f}{\partial t} &=
\frac{\partial^2 f}{\partial V^2}\! \int\! d{\bf u}\,P({\bf u})\,v^{1-\gamma}
[u^2 v^2-({\bf u}\cdot{\bf v})^2]\\
&\quad +\frac{\partial f}{\partial V}\! \int\! d{\bf u}\,P({\bf u})\, 
g^{1-\gamma}({\bf u}\cdot{\bf g})
\end{split}
\end{equation}
in the leading order. Thus the entire effect of the integration measure is captured by one number, $A$. 

To simplify the first integral on the right-hand side of \eqref{long-mu-2} we write
\begin{eqnarray*}
g^{1-\gamma}({\bf g}\cdot{\bf u}) &=& -v^{1-\gamma} ({\bf v}\cdot{\bf u})\\
&+& v^{-1-\gamma}\left[(1-\gamma)({\bf v}\cdot{\bf u})^2 + v^2u^2\right]
\end{eqnarray*}
where we have kept only the leading and the sub-leading terms.  The integral over the leading term vanishes. Using \eqref{T-3d} and \eqref{VV} we recast Eq.~\eqref{long-mu-2} into
\begin{equation}
\label{kinetic}
\frac{\partial f}{\partial \tau} = v^{-\gamma}
\left[(d-\gamma)\,\frac{\partial f}{\partial v} + v\,\frac{\partial^2 f}{\partial v^2}\right]
\end{equation}
where the modified time variable is given by [we additionally put the factor 
$(r_0\epsilon^{1/\lambda})^{d-1}$  back into the time variable] 
\begin{equation}
\label{tau}
\tau = 2A(r_0\epsilon^{1/\lambda})^{d-1} \rho T t
\end{equation}
Although one cannot \cite{calogero} compute the factor $A$ without knowing the integration measure, it is just a number that can be absorbed into the definition of the time variable to arrive at a universal kinetic equation \eqref{kinetic} that depends only on  the interaction exponent $\lambda$. 

The form of equation \eqref{kinetic} implies that $\tau\sim v^{1+\gamma}$. This suggests a scaling ansatz
\begin{equation}
f=\tau^{-\Lambda d}\Phi(w),\quad w=v\tau^{-\Lambda}\,,\quad\Lambda\equiv(1+\gamma)^{-1}\,.
\label{scaling-3d-form}
\end{equation}
Plugging \eqref{scaling-3d-form} into \eqref{kinetic} we obtain
an ordinary differential equation for $\Phi(w)$ which is solved to yield 
\begin{equation}
\Phi(w)=C\,\exp\!\left\{ -\Lambda^{2}w^{1/\Lambda}\right\}\,,\quad
C = \frac{\Lambda^{2\Lambda d - 1}}{\Omega_d\,\Gamma(\Lambda d)}\,.
\label{scaling-3d-lambda}
\end{equation}
Thus the asymptotic growth, $\langle v\rangle \sim \tau^\Lambda$,  of the average speed and the scaled velocity distribution have universal behaviors, the only parameters that matters are the interaction exponent $\lambda$ and the spatial dimensionality $d$. 

To exemplify the speed growth we note that in three dimensions
\begin{equation*}
\langle v\rangle \sim
\begin{cases}
\tau           & {\rm when}~~  \lambda=\infty ~(\text{hard sphere gas})\\
\tau^{3/4}  &{\rm when}~~ \lambda=12 ~(\text{Lennard-Jones gas})\\
\tau^{1/2}  &{\rm when}~~ \lambda=4 ~~\,(\text{Maxwell molecules})
\end{cases}
\end{equation*}

By definition, the Maxwell molecules (MM) interaction \cite{MM_comment} leads to the collision integral that is independent on the relative velocity. Equation \eqref{B-3d-gen} shows that the MM interaction is characterized by $\gamma=1$, so
the interaction exponent is given by $\lambda=2(d-1)$. Interestingly, for the MM particle-atoms interaction, the average velocity experiences {\it standard diffusion} and the scaled particle velocity distribution is Gaussian.  

Let us now estimate the range of the validity of the above results if the particle mass $m$ is small but finite: $0<m\ll 1$. For a while, the evolution follows the zero-mass limit, but eventually the particle equilibrates with the background. The crossover to this regime occurs when the particle velocity becomes of the order of
\begin{equation*}
v_c\sim \sqrt{\frac{T}{m}}
\end{equation*}
In the earlier regime, $t<t_c$, we have $\langle v\rangle \sim \tau^{1/(1+\gamma)}$. The crossover time $t_c$ is therefore estimated from 
\begin{equation*}
(r_0\epsilon^{1/\lambda})^{d-1} \rho T t_c \sim \left(\frac{T}{m}\right)^{\frac{1+\gamma}{2}}
\end{equation*}
that is,
\begin{equation}
\label{t_c}
t_c \sim (r_0\epsilon^{1/\lambda})^{-(d-1)} \rho^{-1}\, T^{-\frac{1-\gamma}{2}}\, m^{-\frac{1+\gamma}{2}}
\end{equation}
The dependence of the crossover time $t_c$ on the gas density and the mass of the particle is easy to appreciate. On the other hand, the dependence of the crossover time on the gas temperature is a bit surprising: 
\begin{enumerate}
\item When $\gamma<1$, that is $\lambda > 2(d-1)$ implying that the potential is harder than the MM potential, the crossover time decreases as the temperature increases. 
\item When $\gamma>1$, that is $\lambda < 2(d-1)$ implying that the potential is softer than the MM potential, the crossover time increases as the temperature increases. 
\end{enumerate}

\section{Displacement of the impurity}
\label{displacement}

We now turn to the spatial behavior of the impurity. We begin with a heuristic analysis. In one dimension, the mean-free path is $\ell=\rho^{-1}$, the average speed grows as $\rho T t$ [see Eq.~\eqref{B-1d-scaling}], and hence the time interval between collisions is $\Delta t\sim \rho^{-1}/(\rho T t)$. This leads to an estimate for the total number of collisions during the time interval $(0,t)$ 
\begin{equation}
\mathcal{N}\sim \frac{t}{\Delta t}\sim \frac{T t^2}{\ell^2}
\end{equation}
The standard random walk argument tells us that a typical displacement of the particle is given by
\begin{equation}
\label{xt-heuristic}
x_{\rm typ}\sim \ell \sqrt{\mathcal{N}}\sim \sqrt{T}\, t
\end{equation}
Hence the displacement exhibits a ballistic, $x\sim t$, rather than diffusive growth with time. Another unexpected feature of the growth law \eqref{xt-heuristic} is that the gas density $\rho$ does not affect the asymptotic.

The situation remains the same for an arbitrary dimension $d$ and an arbitrary interaction. Consider first the hard-sphere interaction. The mean-free path is  $\ell\sim (\rho a^{d-1})^{-1}$ and the average speed is $v\sim \rho a^{d-1} Tt$, see Eq.~\eqref{final_v}. Proceeding as in the one-dimensional case we find
\begin{equation*}
\mathcal{N}\sim \frac{t}{\Delta t}\sim \frac{T t^2}{\ell^2}
\end{equation*}
and therefore 
\begin{equation}
\label{displ}
r_{\rm typ}\sim \ell\sqrt{\mathcal{N}}\sim \sqrt{T}\, t
\end{equation}
The striking feature of this growth law is that the displacement is asymptotically independent on the density of atoms and their size. 

If the particle mass $m$ is small but finite, $0<m\ll 1$, the growth law \eqref{displ} holds up to the crossover time $t_c$ when the displacement becomes of the order of 
\begin{equation}
\label{displ:at}
r_c\sim (r_0\epsilon^{1/\lambda})^{-(d-1)} \rho^{-1}\, T^{\frac{\gamma}{2}}\, m^{-\frac{1+\gamma}{2}}
\end{equation}
while for $t>t_c$ the ballistic growth \eqref{displ} switches to the diffusive growth
\begin{equation}
\label{displ:above}
r_{\rm typ}\sim r_c \sqrt{t/t_c}
\end{equation}

The above heuristic argument can be extended to the case when the particle-atoms interaction is described by a potential. At any time, the model is close to the hard-sphere case with effective radius of the order of $r_*$. But since the displacement growth \eqref{displ} is independent on $a$ in the hard-sphere case, it will be independent on $r_*$ at any given moment, and generally independent on the parameters of the interaction potential \eqref{Ur}. Thus the displacement obeys the same growth law \eqref{displ} independently on $\lambda$ and $d$. 

We now turn from heuristics to exact analyses. To determine the second moment of the spatial distribution we first express it through the velocity correlation function
\begin{eqnarray}
\label{x2-av}
\langle x^2(t)\rangle &=& \int_0^t dt_1  \int_0^t dt_2\,\langle v(t_1)v(t_2)\rangle
\nonumber\\
&=& 2\int_0^t dt_1  \int_{t_1}^t dt_2\,\langle v(t_1)v(t_2)\rangle
\end{eqnarray}
To evaluate $\langle v(t_1)v(t_2)\rangle$ let us consider the impurity particle that starts at the origin with velocity equal to zero (initial conditions are actually irrelevant as we are interested in the long time behavior however this particular choice makes the computation more compact). 
In this case the probability distribution for $v_1=v(t_1)$ is given by Eq.~\eqref{scaling-1d}. To determine the velocity distribution of $v_2=v(t_2)$ we must use $v_1$ as the initial condition. The corresponding distribution function (i.e. the conditional probability) $f(v_2, t_2| v_1, t_1)$ satisfies a kinetic equation which is different from \eqref{B-1d-scaling} as the derivation of the latter assumes that the distribution function is symmetric, $f(v)=f(-v)$. Generally we write

\begin{equation*}
f(v)=
\begin{cases}
f_+(v)& v>0\\
f_-(-v)& v<0
\end{cases}
\end{equation*}
and then proceed as in Sect.~\ref{1D} to yield 
\begin{subequations}
\begin{align}
\frac{\partial f_+}{\partial t} &=2\rho T
\left[\frac{\partial f_-}{\partial v}+v\,\frac{\partial^2 f_-}{\partial v^2}\right]
-\rho v(f_+-f_-)
\label{F+}\\
\frac{\partial f_-}{\partial t} &=2\rho T
\left[\frac{\partial f_+}{\partial v}+v\,\frac{\partial^2 f_+}{\partial v^2}\right]
+\rho v(f_+-f_-)
\label{F-}
\end{align}
\end{subequations}
Subtracting \eqref{F-} from \eqref{F+} we see that the anti-symmetric part 
\begin{equation}
\label{anti-symmetric}
\phi(v)=f_+(v)-f_-(v)
\end{equation}
 satisfies a closed equation
\begin{equation}
\label{1d-anti}
\frac{\partial \phi}{\partial t} = -2\rho T
\left[\frac{\partial \phi}{\partial v}+v\,\frac{\partial^2 \phi}{\partial v^2}\right]
-2\rho v \phi
\end{equation}
(while for the symmetric part $\psi(v)=f_+(v)+f_-(v)$, we recover Eq.~\eqref{B-1d-scaling}).
The initial condition is
\begin{equation}
\phi(v,t=t_1) = \delta(v-v_1)
\end{equation}
and the boundary condition, which follows immediately from the definition Eq.~\eqref{anti-symmetric}, is
\begin{equation}
\label{1d-anti-BC}
\phi(v=0,t) = 0
\end{equation}
The initial-boundary value problem \eqref{1d-anti}--\eqref{1d-anti-BC} is non-trivial, yet in the interesting long time limit the governing equation \eqref{1d-anti} simplifies to $\frac{\partial \phi}{\partial t} = -2\rho v \phi$ (since $v\gg \sqrt{T}$), or equivalently $\frac{\partial \phi}{\partial \tau} = -v \phi/T$. Therefore 
\begin{equation}
\label{1d-phi-sol}
\phi(v,t| v_1, t_1) = \delta(v - v_1)\,e^{-v (\tau-\tau_1)/T}
\end{equation}

The velocity autocorrelation function can be presented in a rather compact form
\begin{equation}
\label{v12-gen}
\begin{split}
\langle v_1v_2\rangle &= 2 \int_0^\infty dv_1\,v_1 f(1)
\int_{-\infty}^\infty dv_2\,v_2 f(2|1)\\
&= 2 \int_0^\infty dv_1\,v_1f(1)
 \int_0^\infty dv_2\,v_2 \phi(2| 1)
\end{split}
\end{equation}
Note that only the anti-symmetric part of $f(2|1)$ contributes to the 2-points velocity correlation function.
For the higher-points velocity correlation functions both the symmetric and anti-symmetric part appear alternatively. For example the 4-points velocity correlation function can be written as:
\begin{equation*}
\langle v_1v_2v_3v_4\rangle = 2\, \left( \prod_{i=1}^4 \int_0^\infty dv_i\,v_i \right) f(1) \phi(2| 1) \psi(3| 2) \phi(4| 3)
\end{equation*}
where $\psi(v,t|v_2,t_2)$ satisfies Eq.~\eqref{B-1d-scaling} with the symmetric initial condition $\psi(v,t=t_2)=\delta(v-v_2)+\delta(v+v_2)$.

Substituting into \eqref{v12-gen} the results for $f(1)\equiv f(v_1,t_1)$  and $\phi(2|1)\equiv \phi(v_2, t_2|v_1,t_1)$ [Eqs.~\eqref{scaling-1d} and \eqref{1d-phi-sol}] we get
\begin{equation*}
\begin{split}
\langle v_1v_2\rangle &=\int_0^\infty dv_1\,
\frac{v_1^2}{\tau_1}\,\exp\!\left(-v_1\left[\frac{1}{\tau_1}+
\frac{\tau_2-\tau_1}{T}\right]\right)\\
&= \frac{2\tau_1^2}{[1+(\tau_2-\tau_1)\tau_1/T]^3}
\end{split}
\end{equation*}
Note that the equal times velocity autocorrelation function ($t_1=t_2=t$) reduces to
$\langle v^2(t)\rangle = 2\tau^2$. This result directly follows from \eqref{scaling-1d} thereby providing a useful check of the consistency of our calculation of the velocity autocorrelation function. Plugging the velocity autocorrelation function into Eq.~\eqref{x2-av} we obtain
\begin{equation}
\label{x2-av-int2}
\langle x^2\rangle = \frac{1}{\rho^2 T^2}\int_0^\tau d\tau_1\,\tau_1^2  
\int_{\tau_1}^\tau 
\frac{d\tau_2}{[1+(\tau_2-\tau_1)\tau_1/T]^3}
\end{equation}
Computing the integral over $\tau_2$ yields
\begin{equation*}
\langle x^2\rangle = \frac{1}{2\rho^2 T}\int_0^\tau d\tau_1\,\tau_1
\left\{1-\frac{1}{[1+(\tau_2-\tau_1)\tau_1/T]^2}\right\}
\end{equation*}
The first integral $\int d\tau_1\,\tau_1$ provides the leading contribution. Recalling that $\tau=2\rho T t$ we arrive at 
\begin{equation}
\label{x2-asymp}
\langle x^2\rangle \simeq T t^2
\end{equation}
This asymptotically exact result confirms the heuristic prediction \eqref{xt-heuristic}.

One can also compute higher-order velocity correlation functions, e.g. $\langle v_1v_2v_3v_4\rangle$, and use them to compute higher moments of the displacement. For instance,
\begin{equation*}
\langle x^4\rangle = 4!\iiiint\limits_{0<t_1<t_2<t_3<t_4<t} dt_1dt_2dt_3dt_4\,
\langle v_1v_2v_3v_4\rangle
\end{equation*}
These computations are very laborious, so we do not present them; we just mention that using this method we were able to compute the asymptotically exact fourth moment of the displacement,
\begin{equation}
\label{x4-asymp}
\langle x^4\rangle \simeq 5\, T^2\,t^4\,,
\end{equation}
in one dimension. 

Finally we note that the above procedure can be generalized to higher dimensions. Even in the case of the hard-sphere particle-atom interaction, however, the explicit computations are quite unwieldy. 

\section{Velocity-Position Distribution}
\label{VPD}

The calculations of the moments of the displacement, e.g. the derivation of equation \eqref{x4-asymp},  through the velocity correlation functions are very cumbersome. It seems hardly possible to succeed in deriving the next moment,
\begin{equation}
\label{x6-asymp}
\langle x^6\rangle \simeq 61\, T^3\,t^6,
\end{equation}
relying on the velocity correlation functions. 

Therefore we employ different procedures that utilize a Boltzmann equation for the velocity-position distribution $f({\bf r}, {\bf v}, t)$. This joint distribution function provides a complete description of the evolution of the impurity particle. Recall that in studying the velocity distribution function we relied on a shorten description for the velocity distribution function $f({\bf v}, t)$. In studying the displacement one would also like to use a governing equation for the density function $N({\bf r}, t)$ as a starting point. Unfortunately, there is no closed equation for the density function $N({\bf r}, t)$. 

In the one-dimensional setting, the governing kinetic equation for the joint distribution $f(x,v,t)$ reads
\begin{equation}
\label{kinetic-1d}
\frac{\partial f}{\partial t}+v\,\frac{\partial f}{\partial x}
=2\rho T\left(\frac{\partial f}{\partial v}+v\,\frac{\partial^2 f}{\partial v^2}\right)
\end{equation}
The left-hand side of this equation is exact, yet Eq.~\eqref{kinetic-1d} is already a simplified version of the Boltzmann equation as the collision term is only asymptotically exact, namely it is appropriate when $v\gg \sqrt{T}$. As we mentioned earlier there is no closed equation for the density function, $N({\bf r}, t)$. 
If one tries to integrate the kinetic equation \eqref{kinetic-1d} over $v$, the convective term leads to a current term, i.e. $\frac{\partial}{\partial x} \int dv\,v f(v,x,t) \equiv \frac{\partial}{\partial x} J(x,t)$, so the density is coupled to the current. One can then deduce from \eqref{kinetic-1d} an equation for the current, but it will involve the second moment $\int dv\,v^2 f(v,x,t)$. This procedure leads to an infinite hierarchy which seems intractable as (essentially) all infinite hierarchies. 

The kinetic equation \eqref{kinetic-1d} is a linear partial differential equation with two coefficients depending linearly on the velocity $v$. The most difficult term in Eq.~\eqref{kinetic-1d}, namely the convective term $(v\nabla)f$, can be further simplified in the long time limit when $v\gg \sqrt{T}$. Indeed, since the particle speed grows (on average) with a constant rate, the particle experiences numerous collisions during a time interval when its speed is almost constant. Then the problem is akin to the standard Lorentz gas where the particle undergoes a simple diffusion. The separation between the time scale at which diffusion appears (few collisions) and the time scale at which the particle speed changes appreciably allows us to replace the convective term by the diffusion term of a standard Lorentz gas.
In one dimension, the diffusion coefficient is $D=v/2\rho$, see \cite{book}. In the present case we can use the same formula. Thus Eq.~\eqref{kinetic-1d} becomes
\begin{equation*}
\frac{\partial f}{\partial t}
=2\rho T\left(\frac{\partial f}{\partial v}+v\,\frac{\partial^2 f}{\partial v^2}\right)
+ \frac{v}{2\rho}\,\frac{\partial^2 f}{\partial x^2}
\end{equation*}
As usual, it is convenient to use $\tau=2\rho T t$ as the time variable.  Then the above equation becomes
\begin{equation}
\label{hydrodynamic-1d}
\frac{\partial f}{\partial \tau}
=\frac{\partial f}{\partial v}+v\,\frac{\partial^2 f}{\partial v^2}
+ \frac{v}{4\rho^2 T}\,\frac{\partial^2 f}{\partial x^2}
\end{equation}
In Eq.~\eqref{hydrodynamic-1d} we tacitly assume that $v>0$. This is obvious regarding the last term on the right-hand side as the diffusion coefficient must be positive (the correct expression is $D=|v|/2\rho$). The form $f_v+vf_{vv}$ of the collision term also assumes (see Sect.~\ref{1D}) that $v>0$. There is no need to separately consider negative velocities, it suffices to take into account the reflection symmetry $f(x,v,t) = f(x,-v,t)$.

In the long time limit, the joint distribution function $f(x, v, t)$ should approach the scaling form
\begin{equation}
\label{scaling-f}
f(x, v, t) \simeq \frac{1}{4 x_{*} v_{*}}\, F(X, V),\quad X=\frac{x}{x_*},\quad V=\frac{v}{v_*}
\end{equation}
where $x_*=\sqrt{T}t$ and $v_*=\tau$. 
The reflection symmetry with respect of the velocity and the displacement \cite{sym} allows us to limit ourself to the quadrant $V>0, X>0$.

By inserting \eqref{scaling-f} into \eqref{hydrodynamic-1d} we obtain
\begin{equation}
\label{F-eq}
2F+X\,\frac{\partial F}{\partial X}+V\,\frac{\partial F}{\partial V}
+\frac{\partial F}{\partial V}+V\,\frac{\partial^2 F}{\partial V^2}
+V\,\frac{\partial^2 F}{\partial X^2}=0
\end{equation}

The normalization condition 
\begin{equation*}
\int_{-\infty}^\infty dx \int_{-\infty}^\infty dv\,f(x, v, t)=1
\end{equation*}
can be re-written as
\begin{equation}
\int_0^\infty dX \int_0^\infty dV\,F(X, V)=1
\end{equation}
This explains the factor $1/4$ in the scaling ansatz \eqref{scaling-f}.

\begin{figure}
\begin{tabular}{cc}
\includegraphics[width=0.9\columnwidth]{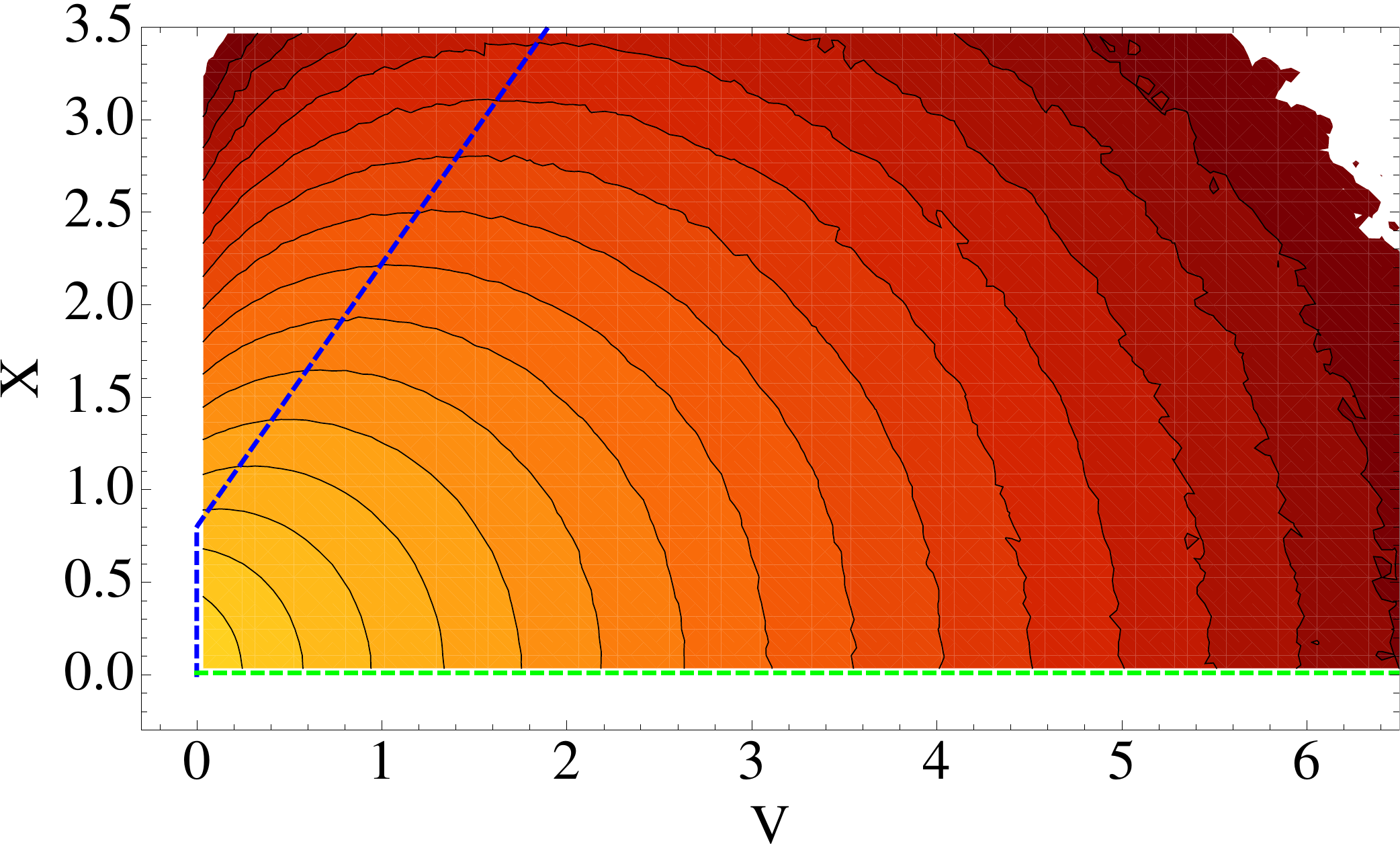} \\ 
\includegraphics[width=0.9\columnwidth]{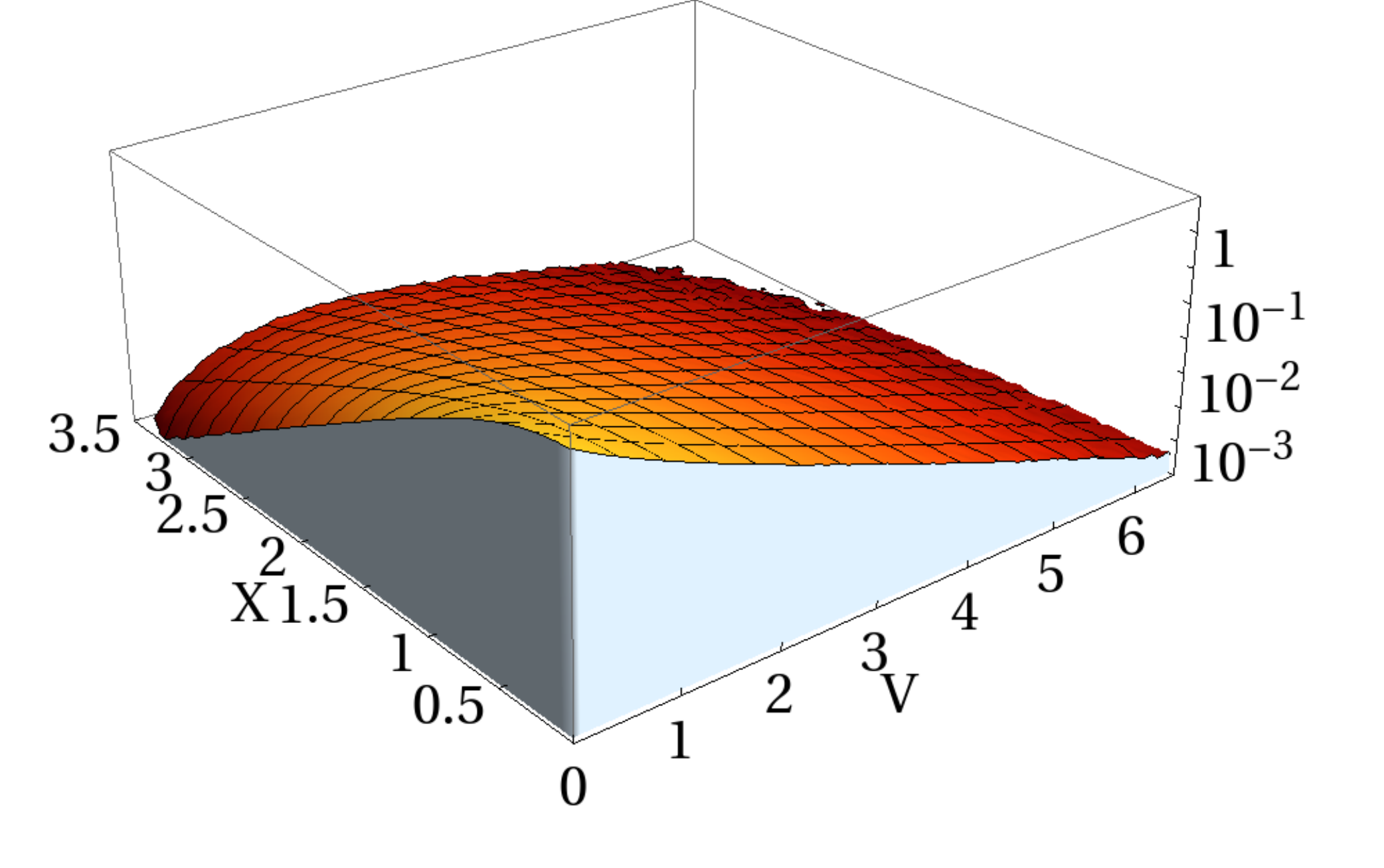} 
\end{tabular}
\caption{Shown are simulation results (see Sect.~\ref{numerics}). Contour-plot (top panel) and 3D-plot (bottom panel) of the distribution function $F(X,V)$ for a gas of hard spheres in one dimension. For any given velocity (position) the green (blue) dashed line shows the value of the position (velocity) for which the probability distribution has a maximum.} 
\label{joint}
\end{figure}

In higher dimensions, we limit ourselves to the case of the hard-core particle-atoms interaction. Then the governing kinetic equation reads 
\begin{equation}
\label{kinetic-d}
\frac{\partial f}{\partial t}+{\bf v}\cdot\frac{\partial f}{\partial {\bf r}}
=2 a^{d-1}A\rho T\left(d\,\frac{\partial f}{\partial v}+v\,\frac{\partial^2 f}{\partial v^2}\right) + D\nabla^2 f
\end{equation}
Equation \eqref{kinetic-d} is again asymptotically exact in the large time limit when the typical particle velocity greatly exceeds the thermal velocity, $v\gg \sqrt{T}$. In this limit, the collision term simplifies to the first term on the right-hand side of Eq.~\eqref{kinetic-d} and the convective term $({\bf v}\cdot {\bf \nabla})f$ can be replaced by the diffusion term $-D\nabla^2 f$ as the transport is asymptotically diffusion with velocity-dependent diffusion coefficient. More precisely, the diffusion coefficient is given by  \cite{diff_general}
\begin{equation}
\label{D_general}
D = \frac{v}{2dA a^{d-1} \rho}
\end{equation}
with the amplitude $A$ known in the case of the hard-core interaction, see \eqref{Ad}. Using again  the modified time variable is $\tau$ [which for hard-sphere particle-atom interaction is given by $\tau = 2A a^{d-1} \rho T t$, see \eqref{final_v}], and taking into account the spatial isotropy we recast \eqref{kinetic-d} into
\begin{eqnarray}
\label{hydrodynamic-d}
\frac{\partial f}{\partial \tau}
&=& d\,\frac{\partial f}{\partial v}+v\,\frac{\partial^2 f}{\partial v^2}\nonumber\\
&+& \frac{v}{d(2A a^{d-1}\rho)^2 T}\left(\frac{\partial^2 f}{\partial r^2} 
+ \frac{d-1}{r}\,\frac{\partial f}{\partial r}\right)
\end{eqnarray}

A solution to Eq.~\eqref{hydrodynamic-d} approaches a scaling form
\begin{equation}
\label{scaling-fd}
f(r,v,t)=(\Omega_d)^{-2} \left(\tau \sqrt{T}\,t\right)^{-d}\, F(V, R)
\end{equation}
with scaled spatial and velocity variables
\begin{equation}
\label{scaling-RV}
R = \frac{r}{\sqrt{T}\,t}\,, \quad V = \frac{v}{\tau}
\end{equation}
With the choice \eqref{scaling-fd} of the scaling form, the normalization requirement 
\begin{equation*}
\int_0^\infty \Omega_d r^{d-1} dr 
\int_0^\infty \Omega_d v^{d-1} dv\, f({\bf r}, {\bf v}, t) = 1
\end{equation*}
becomes
\begin{equation}
\label{norm-d}
\int_0^\infty dR\int_0^\infty dV\, R^{d-1} V^{d-1} F(R,V) =1
\end{equation}
Using \eqref{scaling-fd}--\eqref{scaling-RV} we transform \eqref{hydrodynamic-d} into 
\begin{eqnarray}
\label{FRV}
2dF\!\!&+\!\!&RF_R+VF_V+ dF_V+VF_{VV}\nonumber\\
 &+\!\!&\frac{V}{d}\!\left(\frac{d-1}{R}\,F_R+F_{RR}\right)=0
\end{eqnarray}
This is a linear elliptic (recall that $R>0, V>0$) partial-differential equation. Despite of linearity, Eq.~\eqref{FRV} is difficult since the coefficients in front of derivatives in Eq.~\eqref{FRV} vary with $V$ and $R$. 

We treat above equations by using different techniques. The standard technique relying on the Laplace and Fourier transforms is the most powerful. In Sect.~\ref{full_solution} we derive the major result for the scaled joint distribution of the impurity particle in the hard-sphere gas: 
\begin{equation}
\label{joint_d}
F(R,V) = \frac{C_d}{\Gamma(d)}\int d{\bf s}\, 
e^{-i \sqrt{d}\, {\bf s}\cdot {\bf R}-Vs\coth s}\left(\frac{s}{\sinh s}\right)^d
\end{equation}
Further, the scaled density distribution reads
\begin{equation}
\label{density_d}
N(R) = C_d\int d{\bf s}\,\frac{e^{-i \sqrt{d}\, {\bf s}\cdot {\bf R}}}{(\cosh s)^d}\,, \quad 
C_d = \frac{d^{d/2} \Omega_d}{(2\pi)^d}
\end{equation}
In particular, in one dimension
\begin{equation}
\label{NX}
N(X) = \frac{1}{\cosh R_1}\,,\quad R_1=\frac{\pi}{2}\, X
\end{equation}
while in three dimensions the density is
\begin{equation}
\label{NR3}
N(R) = \frac{3\sqrt{3}}{8}\, \frac{(4R_3^2+\pi^2)\tanh R_3-8R_3}{R_3 \cosh R_3}\,,
\quad R_3=\frac{\pi\sqrt{3}}{2}\, R
\end{equation}

First, however, we describe an approach based on the direct computing of the moments and guessing from them the spatial distribution. 

\section{Moments}
\label{moment_approach}

The moment approach deals with the moments of the joint distribution rather than with the joint distribution itself. The moment approach has been used in kinetic theory throughout its history (see e.g. \cite{MAX,TM80}) as the governing equations are very complicated and seldom tractable. The moment approach has also been applied~\cite{poland,italy} to the Fermi's acceleration mechanism. For instance, in Refs.~\cite{poland} the authors computed the moments $\langle v^n\rangle$ for small $n$, guessed the answer [namely \eqref{vn-av}] for an arbitrary $n$, showed that the guess is correct, and observed that the exponential velocity distribution has exactly the same moments. Generally if one succeeds in computing the moments, one still has to recover the distribution that has such moments. This is not rigorous as at best we have infinitely many {\em integer} moments (or only even integer moments as in examples below) and we want to restore the entire distribution function. If the distribution function is analytic (the fact which is usually unknown, but believed to be correct), the distribution function can be uniquely determined by (infinitely many) integer moments, so restoring such function is a technical problem. 

Another problem is that since the number of moments is infinite, it is usually impossible to compute them all. Having computed a few moments one can try to guess the rest and to check the conjecture using computer-assisted {\em exact} calculations. We have succeeded in guessing all even moments of the spatial displacement in one and two dimensions, and in reading off the density in one dimension. The moment approach is therefore not really systematic and it involves a guess work. The strength of the moment approach is that one can easily compute the basic moments, e.g. even moments of the displacement $\langle R^2\rangle, \langle R^4\rangle, \langle R^6\rangle$, etc., or mixed moments like $\langle R^2 V^2\rangle$, and arrive at important conclusions (like the existence of correlations between the velocity and the spatial displacement manifested by relation $\langle R^2 V^2\rangle \ne \langle R^2\rangle \langle V^2\rangle$). 

In our problem we eventually derived more comprehensive results using standard techniques (see Sect.~\ref{full_solution}). Still, the moment approach has a future. Indeed it is more powerful nowadays than it ever was as the tedious calculations of the moments can be exactly performed by a computer and if the resulting moments admit a simple expression through well-known sequences, there is a good chance to extract such an expression by using \textit{The On-Line Encyclopedia of Integer Sequences} \cite{enciclopedia}. Since the moment approach is rarely used, we illustrate it here as in our situation where the moment approach clearly gives highly non-trivial results. We begin with the one-dimensional setting.

\subsection{One Dimension}

In this subsection we will present a very strong evidence in favor of the announced result \eqref{NX}
for the spatial distribution. To establish \eqref{NX}, we turn \eqref{F-eq} into an infinite set of relations 
\begin{equation}
\label{Mij-eq}
(i+j)M_{i,j} = j^2 M_{i,j-1}+i(i-1)M_{i-2,j+1}
\end{equation}
for the moments
\begin{equation}
\label{Mij-def}
M_{i,j}= \int_{0}^\infty \int_{0}^\infty dX\,dV\,X^i V^j F(X,V)
\end{equation}
The relation \eqref{Mij-eq} is valid for all $i\geq 2, j\geq 0$. 

\begin{figure}
\includegraphics[width=0.9\columnwidth]{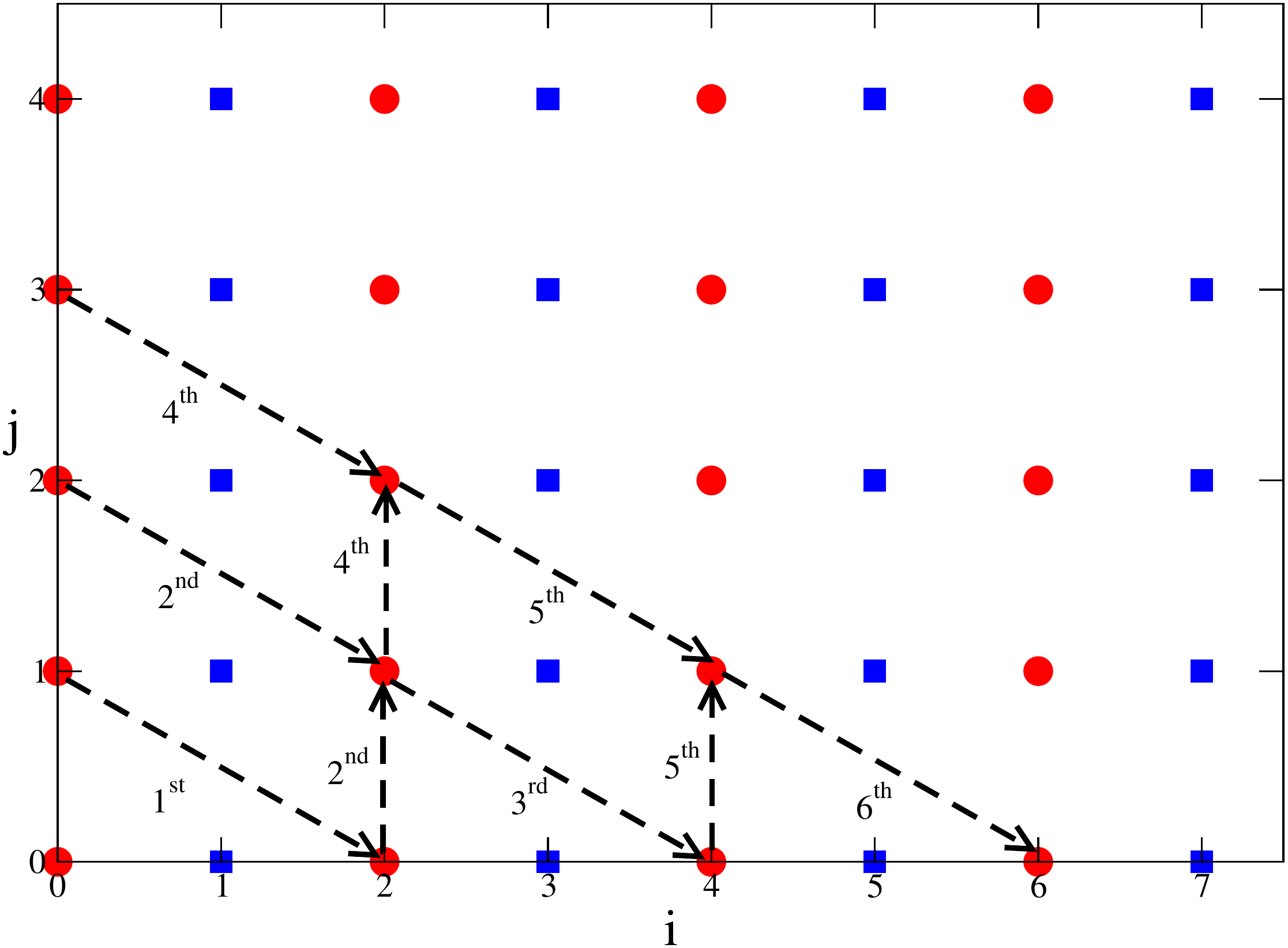}
\caption{Schematic representation of how Eqs.~\eqref{Mij-eq} can be iteratively used to calculate all the moments $M_{i,j}=\langle R^i V^j\rangle$ with $i=even$ (red circles). The moments $M_{0,j}$ are known for all $j\geq 0$. At the first step the known value of $M_{0,1}$ allows us to calculate $M_{2,0}$. At the second step the already known $M_{2,0}$ and $M_{0,2}$ are used to calculate $M_{2,1}$, see \eqref{21}. At the third step we compute $M_{4,0}$ through $M_{2,1}$. 
The moments $M_{i,j}$  with $i=odd$ (blue squares) cannot be calculated using this approach.}
\label{recurrence}
\end{figure}

Using \eqref{Mij-eq} one can compute moments with small indexes; for instance, one can establish \eqref{x2-asymp}--\eqref{x6-asymp}. Figure~\eqref{recurrence} illustrates the structure of the quasi-recurrent equation \eqref{Mij-eq} and the procedure to calculate the first few spatial moments. One finds that $\langle X^{2n}\rangle=M_{2n,0}$ can be expressed as a weighted sum of $M_{0,1},\ldots,M_{0,n}$. This sum is then computed using the identity
\begin{equation}
\label{M0j}
M_{0,j} = \langle V^j\rangle = \int_0^\infty dV\,e^{-V}V^j=j!
\end{equation}

We now demonstrate this in practice. Specializing \eqref{Mij-eq} to $(i,j)=(2,0)$ gives $M_{2,0} = M_{0,1}=1$ which is identical to Eq.~\eqref{x2-asymp}. Specializing \eqref{Mij-eq} to $(i,j)=(2,1)$ yields
\begin{equation}
\label{21}
3M_{2,1} = M_{2,0} + 2M_{0,2}
\end{equation}
Taking then $(i,j)=(4,0)$ we obtain $M_{4,0}  = 3M_{2,1}$, or
\begin{equation}
\label{40}
M_{4,0} = M_{0,1}+2M_{0,2} = 5
\end{equation}
which is equivalent to \eqref{x4-asymp}. Further, specializing \eqref{Mij-eq} to 
$(i,j)=(6,0), (4,1), (2,2)$ and using \eqref{40} we obtain 
\begin{equation}
\label{60}
M_{6,0} = 5M_{0,1}+10M_{0,2} + 6M_{0,3} = 61
\end{equation}
which proves \eqref{x6-asymp}. The fact that we have been able to reproduce the values of the spatial moments calculated using the velocity correlation functions (Eqs.~\eqref{x2-asymp}--\eqref{x6-asymp}) supports the claim that the replacement of the convection term by the diffusion term in Eq.~\eqref{kinetic-1d} is asymptotically exact.

The computed even moments $\langle X^{2n}\rangle$ are all integers which look familiar; indeed, up to the sign they are the Euler's numbers
\begin{equation}
\label{X-even}
\langle X^{2n}\rangle =(-1)^n E_{2n}
\end{equation}
The Euler's numbers $E_n$ appear in numerous combinatorial problems, as well as in number theory, topology, etc. The Euler's numbers are defined by the Taylor series
\begin{equation}
\label{E-def}
\frac{1}{\cosh(y)}=\sum_{n\geq 0}\frac{E_n y^n}{n!}
\end{equation}
Note that all the odd-indexed Euler numbers are equal to zero, while the even-indexed Euler number have alternating signs.

The evidence in the exactness of \eqref{X-even} is overwhelming --- using {\it Mathematica}, we verified \eqref{X-even} for all even moments up to 
$\langle X^{1000}\rangle$. 

To establish \eqref{NX} we start by extending the range of $X$ to the whole axis and calculate the Fourier transform of $N_{\rm sym}(X)=\tfrac{1}{2}N(|X|)$:
\begin{eqnarray}
\label{cosh2}
\widehat{N}_{\rm sym}(s) & = & \int_{-\infty}^\infty dX\,e^{-isX} N_{\rm sym}(X)\nonumber\\
   & = & \sum_{n\geq 0} \frac{(-1)^n s^{2n} \langle X^{2n}\rangle}{(2n)!}\nonumber\\
   & = & \sum_{n\geq 0} \frac{s^{2n} E_{2n}}{(2n)!}= \frac{1}{\cosh s}
\end{eqnarray}
where on the first step we have expanded $e^{-isX}$ and taken into account that $N_{\rm sym}(X)$ is an even function of $X$, while on the second and third steps we have used \eqref{X-even} and \eqref{E-def}, respectively.  
Since  
\begin{equation}
\label{cosh1}
\frac{1}{2}\int_{-\infty}^\infty dX\,\frac{e^{-isX}}{\cosh(\pi X/2)}= \frac{1}{\cosh s}
\end{equation}
we conclude that $N_{\rm sym}(X)=1/[2 \cosh{(\pi X /2 )}]$ which is equivalent to Eq.~\eqref{NX}.

The moment relations \eqref{Mij-eq} have helped us to determine all even moments $\langle X^{2n}\rangle$, yet they do not allow one to determine even the simplest odd moment $\langle X\rangle$. Using the spatial density \eqref{NX}, however, we can compute this moment (more precisely it is equal to $\langle |X|\rangle$ and it represents the dimensionless average displacement):
\begin{equation*}
\langle |X|\rangle = \int_0^\infty dX\,\frac{X}{\cosh(\pi X/2)} = \frac{8G}{\pi^2}
\end{equation*}
where $G$ is the Catalan constant
\begin{equation*}
G = \frac{1}{1^2}-\frac{1}{3^2}+\frac{1}{5^2}-\frac{1}{7^2}+\cdots 
= 0.915 965 594\ldots
\end{equation*}
Hence the average displacement is given by
\begin{equation*}
\langle |x|\rangle =  \frac{8G}{\pi^2}\,\sqrt{T}\,t
\end{equation*}
Similarly, one can compute an arbitrary odd moment
\begin{equation*}
\langle |X|^{2k-1}\rangle = \frac{2^{2k+1} (2k-1)!}{\pi^{2k}}\,\sum_{m\geq 0}\frac{(-1)^m}{(2m+1)^{2k}}
\end{equation*}

We can establish some qualitative and quantitative features of the joint distribution without having its analytical expression. For instance, if the joint distribution has allowed the factorization, that is if it had the form $N(X)F(V)$, then the moments would satisfy $\langle |X|^i ~ |V|^j\rangle = \langle |X|^i \rangle \langle |V|^j\rangle$. This is not so, e.g.
\begin{equation*}
\frac{\langle X^2 V\rangle}{\langle X^2 \rangle \langle V \rangle}
= \frac{5}{3}\,, \quad 
\frac{\langle X^2 V^2\rangle}{\langle X^2 \rangle \langle V^2 \rangle}
= \frac{7}{3}\,, \quad 
\frac{\langle X^4 V^2\rangle}{\langle X^4 \rangle \langle V^2 \rangle}
= \frac{331}{75}
\end{equation*}
etc. Qualitatively, these results are not surprising --- the larger separation from the starting position, the larger (on average) the speed of the particle is expected to be. Mathematically, this implies an inequality
\begin{equation}
\label{ineq-ij}
\frac{\langle |X|^i ~ |V|^j\rangle}{ \langle |X|^i \rangle \langle |V|^j\rangle} >1
\end{equation}
for all $i,j>0$. This inequality is indeed obeyed in all instances where we were able to compute the moments, for instance when both indexes are sufficiently small. Using Eqs.~\eqref{Mij-eq} we have also computed a few infinite series, e.g.  
\begin{equation}
\label{series}
\begin{split}
\frac{\langle X^{2n}  |V|\rangle}{\langle X^{2n} \rangle \langle |V|\rangle}
&= \frac{1}{2n+1}\,\frac{|E_{2n+2}|}{|E_{2n}|} >1\\
\frac{\langle X^{2}  |V|^j\rangle}{\langle X^{2} \rangle \langle |V|^j\rangle}
&=1 + \frac{2}{3}\,j\\
\frac{\langle X^{4}  |V|^j\rangle}{\langle X^4 \rangle \langle |V|^j\rangle}
&=1 + \frac{88}{75}\,j + \frac{4}{15}\,j^2 \\
\frac{\langle X^{6}  |V|^j\rangle}{\langle X^6 \rangle \langle |V|^j\rangle}
&=1 + \frac{794}{549}\,j + \frac{116}{183}\,j^2+ \frac{40}{549}\,j^3 
\end{split}
\end{equation}
Thus in these cases the inequality \eqref{ineq-ij} is valid. 

The correlation between the velocity and the displacement of the particle shows that the knowledge of the velocity distribution $F(V)$ and the density $N(X)$ provides a limited information about the characteristics of the particle --- the joint distribution function $F(X,V)$ is needed to provide a complete (in the realm of kinetic theory) description. 

\subsection{Higher Dimensions} 

The normalization condition \eqref{norm-d} suggests to define the moments via
\begin{equation}
\label{moments}
M_{i,j}=\int_{0}^{\infty}dR\int_{0}^{\infty}dV\, R^{i+d-1}V^{j+d-1}F(V,R)
\end{equation}
Multiplying equation \eqref{FRV} by $R^{i+d-1}V^{j+d-1}$ and integrating we arrive at the moment relations 
\begin{eqnarray}
\label{Mij-d}
(i+j)M_{i,j} &=& j(j+d-1) M_{i,j-1}\nonumber \\
&+&\frac{i(i+d-2)}{d}\,M_{i-2,j+1}
\end{eqnarray}

We can now proceed as in the one-dimensional case. Namely using relations \eqref{Mij-d}, we can in principle exactly compute any moment $\langle R^{2n}\rangle=M_{2n,0}$ by expressing it as a weighted sum of $M_{0,1},\ldots,M_{0,n}$. Then we use the known expression for $M_{0,j}$
\begin{equation}
\label{M0j-d}
M_{0,j} = \langle V^j\rangle
 = \int_0^\infty dV\,\frac{e^{-V}}{\Gamma(d)}\,V^{j+d-1}
 = \frac{\Gamma(j+d)}{\Gamma(d)}
\end{equation}
which is computed with the help of Eq.~\eqref{scaling-vd}. This procedure gives
\begin{subequations}
\begin{align}
\langle R^2\rangle &= d
 \label{R2:d}\\
\langle R^4\rangle &= (d+2)(d+\tfrac{2}{3})
 \label{R4:d}\\
\langle R^6\rangle &= d^{-1}(d+2)(d+4)(d^2+2d+\tfrac{16}{15})
 \label{R6:d}
\end{align}
\end{subequations}

Using {\it Mathematica}, we have computed the moments $\langle R^{2n} \rangle = M_{2n,0}$ up to $\langle R^{1000} \rangle$ in two and three dimensions. A few of these even-indexed moments are listed in Table I. In contrast to one-dimensional results (also presented in Table I), the moments are no longer integer; apparently \cite{integer} they are non-integer for all (even) $n\geq 4$. 

\begin{table}[t]
\begin{tabular}{|c|c|c|c|}
\hline
$n$ & $1d$ & $2d$ & $3d$\\
\hline
$0$   & $1$                       & $1$                                                 & $1$ \\
$2$   & $1$                       & $2$                                                 & $3$ \\
$4$   & $5$                       & $\frac{32}{3}$                                  & $\frac{55}{3}$\\
$6$   & $61$                     & $\frac{544}{5}$                                & $\frac{1687}{9}$\\
$8$   & $1385$                 & $\frac{63\,488}{35}$                        & $\frac{8651}{3}$\\
$10$ & $50\,521$             & $\frac{2\,830\,336}{63}$                  & $\frac{5\,047\,691}{81}$\\
$12$ & $2\,702\,765$       & $\frac{357\,892\,096}{231}$            & $\frac{437\,804\,783}{243}$\\
$14$ & $199\,360\,981$   & $\frac{30\,460\,116\,992}{429}$      & $\frac{16\,325\,727\,605}{243}$\\
$16$ & $19\,391\,512\,145$ & $\frac{26\,862\,763\,900\,928}{6435}$& $\frac{6\,868\,768\,364\,827}{2187}$\\
\hline
\end{tabular}
\caption{The moments $\langle R^n\rangle$ in one, two, and three dimensions for small even indexes.}
\label{momentstable}
\end{table}

We tried to identify the sequence  $\langle R^{2n} \rangle = M_{2n,0}$ with known sequences \cite{enciclopedia}. Since most known sequences are integer, one can seek $M_{2n,0}$ as a ratio of integer sequences. In three dimensions one can write $\langle R^{2n} \rangle = \mathcal{M}_n/3^n$. The sequence $\mathcal{M}_n$ is integer, but it does not appear in \cite{enciclopedia}. In two dimensions we were more lucky: Seeking $M_{2n,0}$ as a ratio of integer sequences we arrived at  
\begin{equation}
\langle R^{2n}\rangle=\frac{2^{3n+1}(4^{n+1}-1)}{n+1}
\cdot\frac{n!n!}{(2n)!}|B_{2n+2}|\,,
\label{r2n_2d}
\end{equation}
where $B_{k}$ are the Bernoulli numbers \cite{GKP89}. The evidence in the exactness of \eqref{r2n_2d} is overwhelming (we have checked it up to $n=500$).

\subsection{Tail of the density distribution}

According to our definition of the scaled density distribution $N(R)$, it satisfies
\begin{equation}
\label{N-norm}
\int_0^\infty dR\,R^{d-1}N(R) = 1
\end{equation}
In one dimension, $N=[\cosh(\pi X/2)]^{-1}$, and therefore the tail of the 
distribution is
\begin{equation}
\label{N1-tail}
N \simeq 2\,e^{-\pi X/2}\qquad \text{when}\quad X\to\infty
\end{equation}
This exact asymptotic leads to the conjecture that generally in $d$ dimensions the leading asymptotic is exponential. More precisely, we assume that
\begin{equation}
\label{Nd-tail}
N \simeq C\,R^c\, e^{-\mu R}\qquad \text{when}\quad R\to\infty
\end{equation}
where we have augmented the controlling factor $e^{-\mu R}$ by an algebraic pre-factor $R^c$ and the amplitude $C$. The parameters $\mu, c, C$ are dimensionless, so they can depend only on $d$. 

In principle, the moments 
\begin{equation}
\label{moments:def}
\langle R^{2n}\rangle=\int_0^\infty dR\,R^{2n+d-1}N(R)
\end{equation}
depend on the entire density distribution $N(R)$. In the $n\to\infty$ limit, however, the integral in Eq.~\eqref{moments:def} is chiefly gathered in the tail of the distribution. Hence we can use the ansatz \eqref{Nd-tail}. Plugging it into \eqref{moments:def} we get
\begin{eqnarray}
\label{mom_asymp}
\langle R^{2n}\rangle 
                   &\simeq  & C\int_0^\infty dR\,R^{2n+c+d-1}e^{-\mu R}\nonumber \\
                   &=&          \frac{C}{\mu^{2n+c+d}}\,\Gamma(2n+c+d)
\end{eqnarray}
when $n\gg 1$. 

\begin{figure}
\includegraphics[width=0.9\columnwidth]{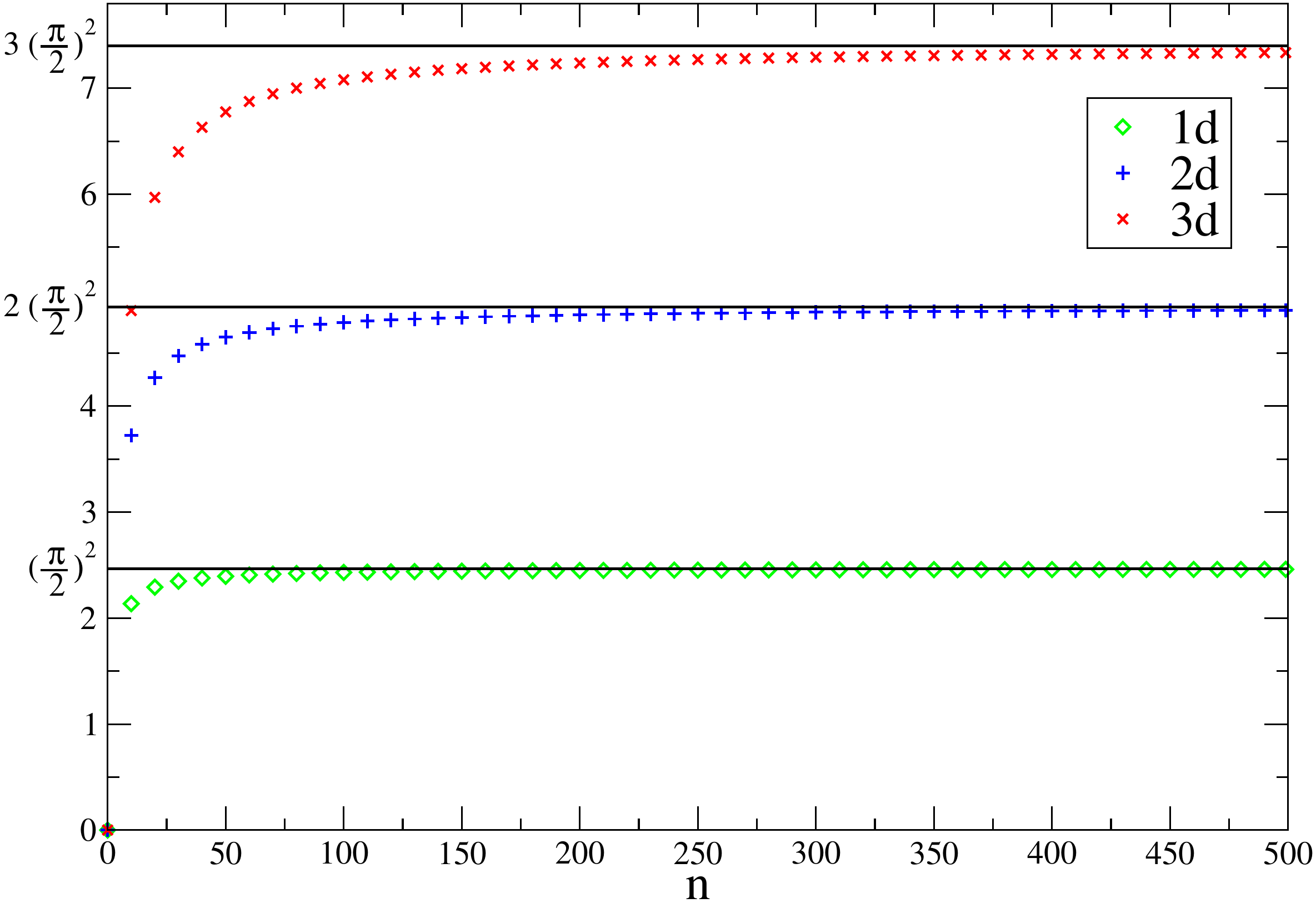}
\caption{Plot of $\frac{(2n)^2 \langle R^{2n}\rangle}{\langle R^{2n+2}\rangle}$ 
vs. $n$ for the hard sphere gas in $d=1,2,3$. Using Eq.~\eqref{ratio_asymptotics} we extract the controlling exponential factor $e^{-\mu_d R}$ of the density profile at large $R$ and we confirm that $\mu _d= \tfrac{1}{2}\pi\sqrt{d}$ in  $d=1,2,3$.}
\label{saturation}
\end{figure}

In two dimensions, Eq.~\eqref{r2n_2d} that yields even moments involves Bernoulli numbers whose asymptotic can be extracted from the celebrated Euler's formula relating Bernoulli's numbers with the values of the zeta function at positive even integers:
\begin{equation}
\label{Euler}
|B_{2k}|=\
\frac{2\,(2k)!}{(2\pi)^{2k}}\,\zeta(2k)\,,\quad \zeta(s)=\sum_{j\geq 1}\frac{1}{j^s}
\end{equation}
Thus we recast  \eqref{r2n_2d} into
\begin{equation*}
\langle R^{2n}\rangle=\frac{2^{3n+1}\big(4^{n+1}-1\big)}{n+1}\,
\frac{n!\,n!}{(2n)!}\,\frac{2\,(2n+2)!}{(2\pi)^{2n+2}}\,\zeta(2n+2)
\end{equation*}
Using Stirling's formula, we simplify the ratio
\begin{equation*}
\frac{n!\,n!}{(2n)!}\simeq \frac{\big(\tfrac{n}{e}\big)^{2n} 2\pi n}
{\big(\tfrac{2n}{e}\big)^{2n} \sqrt{4\pi n}}= \frac{\sqrt{\pi n}}{2^{2n}}
\end{equation*}
We also notice that $\zeta(2n+2)-1\simeq 2^{-2n-2}$, and therefore asymptotically 
$\zeta(2n+2)\simeq 1$ for $n\gg 1$. Thus the moment $\langle R^{2n}\rangle$ approaches to
\begin{equation}
\label{actual}
\langle R^{2n}\rangle\simeq \frac{2^{n+3}}{\pi^{2n+2}}\,\sqrt{\pi n}\, (2n+1)!
\end{equation}
in the $n\to\infty$ limit. On the other hand, in two dimensions the asymptotic prediction \eqref{mom_asymp} based on the ansatz \eqref{Nd-tail} can be re-written in the form
\begin{equation}
\label{emergent}
\langle R^{2n}\rangle\simeq \frac{C}{\mu^{2n+c+2}}\, (2n)^{c}\, (2n+1)!
\end{equation}
where we used the well-known asymptotic \cite{GKP89} 
\begin{equation*}
\frac{\Gamma(m+a)}{\Gamma(m)}\simeq m^a\qquad\text{when}\quad m\to\infty
\end{equation*}
The asymptotics \eqref{actual} and \eqref{emergent} would agree if
\begin{equation*}
\frac{2^{n+3}}{\pi^{2n+2}}\,\sqrt{\pi n} = \frac{C}{\mu^{2n+c+2}}\, (2n)^{c}
\end{equation*}
We get $\mu=\pi/\sqrt{2}$ by matching the dominant exponential factors. Matching then the sub-leading algebraic factors we get $c=1/2$. Matching finally the amplitudes yields $C=2^{5/4}\pi$. Therefore in two dimensions
\begin{equation}
\label{tail2dA}
N\simeq 2^{5/4}\pi\, \sqrt{R}\,\, e^{-\pi R/\sqrt{2}}\quad \text{when}\quad R\to\infty
\end{equation}
The asymptotics in one and two dimensions make plausible that the controlling exponential factor in higher dimensions is
\begin{equation}
\label{Nd-exp-tail}
N\sim \exp\!\left\{-\tfrac{\pi\sqrt{d}}{2} R\right\}
\end{equation}
Thus $N\sim e^{-\mu_3 R}$ with $\mu_3=\tfrac{1}{2}\pi\sqrt{3}\Doteq 2.720699$ in three dimensions. To extract $\mu_3$ we proceed as follows. Using {\it Mathematica}, we have determined the exact values of the moments $\langle R^{2n} \rangle = M_{2n,0}$ up to $\langle R^{1000} \rangle$ in three dimensions. Hence we can compute the ratio of consecutive terms and compare the outcome with the prediction of Eq.~\eqref{mom_asymp}. The latter becomes (in three dimensions)
\begin{equation}
\label{ratio_asymptotics}
\frac{\langle R^{2n} \rangle}{\langle R^{2n+2} \rangle}\simeq
\frac{(\mu_3)^2}{(2n+c+3)(2n+c+4)}
\end{equation}
Thus the quantity $(2n)^2\langle R^{2n} \rangle/\langle R^{2n+2} \rangle$ should converge for $n \rightarrow \infty$ to $(\mu_3)^2=3\pi^2/4 \Doteq 7.402203$.
This is indeed in excellent agreement with our findings (Fig.~\ref{saturation}). 

\subsection{Correlations}

As in the one-dimensional case, both in two and three dimensions there are correlations between the position and the speed of the impurity particle. In this subsection, we present a few results for the three-dimensional case. One can compute $\langle R^i  V^j\rangle$ for even $i$ and arbitrary $j$. For instance
\begin{equation*}
\frac{\langle R^2 V^2\rangle}{\langle R^2 \rangle \langle V^2 \rangle}
= \frac{13}{9}\,, \quad 
\frac{\langle R^2 V^4\rangle}{\langle R^2 \rangle \langle V^4 \rangle}
= \frac{17}{9}\,, \quad 
\frac{\langle R^4 V^2\rangle}{\langle R^4 \rangle \langle V^2 \rangle}
= \frac{991}{495}
\end{equation*}
etc. suggesting again that the inequality
\begin{equation}
\label{ineq-RV}
\frac{\langle R^i  V^j\rangle}{ \langle R^i \rangle \langle V^j\rangle}>1
\end{equation}
is valid for all $i,j>0$. One can compute the left-hand side of Eq.~\eqref{ineq-RV} for arbitrary $j$ and sufficiently small $i$:
\begin{subequations}
\begin{align}
\frac{\langle R^2  V^j\rangle}{ \langle R^2 \rangle \langle V^j\rangle}&
= 1 + \frac{2}{9}\,j
\label{RV-ratios-2}\\
\frac{\langle R^4  V^j\rangle}{ \langle R^4 \rangle \langle V^j\rangle}&
= 1 + \frac{208}{495}\,j+\frac{4}{99}\,j^2
\label{RV-ratios-4}\\
\frac{\langle R^6  V^j\rangle}{ \langle R^6 \rangle \langle V^j\rangle}&
= 1 + \frac{27074}{45549}\,j+\frac{236}{2169}\,j^2+\frac{40}{6507}\,j^3
\label{RV-ratios-6}
\end{align}
\end{subequations}

For instance, let us establish \eqref{RV-ratios-2}. First, we specialize \eqref{Mij-d} to $d=3$ and $i=2$ to yield
\begin{equation}
\label{M2j:3d}
(j+2) M_{2,j}=j(j+2) M_{2,j-1}+2M_{0,j+1}
\end{equation}
Using \eqref{M0j-d} and setting $d=3$ we get $M_{0,j+1}=\tfrac{1}{2}(j+3)!$ and therefore 
\eqref{M2j:3d} becomes
\begin{equation}
\label{M2j:rec}
M_{2,j}=j M_{2,j-1}+(j+3)(j+1)!
\end{equation}
The form of this recurrence suggests to seek $M_{2,j}$ in the form 
$M_{2,j}=j!\,N_j$. This transformation leads to 
\begin{equation}
\label{Nj:rec}
N_j = N_{j-1}+(j+3)(j+1)
\end{equation}
Solving recurrence \eqref{Nj:rec} subject to the `initial' condition $N_0=3$ [this condition ensures that 
$M_{2,0}=\langle R^2\rangle = 3$] we obtain
\begin{equation*}
N_j=3+\sum_{l=1}^j (l+3)(l+1)=\frac{1}{6}(j+1)(j+2)(2j+9)
\end{equation*}
Since $\langle R^2  V^j\rangle = M_{2,j}=j!\,N_j=\tfrac{1}{6}(j+2)!(2j+9)$
and $\langle R^2 \rangle \langle V^j\rangle = 3M_{0,j}=\tfrac{3}{2}(j+2)!$, we have
\begin{equation*}
\frac{\langle R^2  V^j\rangle}{ \langle R^2 \rangle \langle V^j\rangle}
= \frac{\tfrac{1}{6}(j+2)!(2j+9)}{\tfrac{3}{2}(j+2)!} = 
1 + \frac{2}{9}\,j
\end{equation*}
thereby establishing \eqref{RV-ratios-2}. Using similar reasoning we have derived \eqref{RV-ratios-4}--\eqref{RV-ratios-6}, as well as analogous results \eqref{series} in one dimension. 

The ratios \eqref{RV-ratios-2}--\eqref{RV-ratios-6} suggest that
\begin{equation}
\int_{0}^\infty dR\,R^{2+2i} F(R,V) = e^{-V}\,P_i(V)
\end{equation}
with $P_i(V)$ being a polynomial of $V$ of degree $i$. We already know that
$P_0(V)=1/2$ in three dimensions. (Generally $P_0(V)=1/(d-1)!$.)  Using \eqref{RV-ratios-2}--\eqref{RV-ratios-6} we arrive at the following explicit results for the polynomials $P_i(V)$ with $i=1,2,3$:
\begin{equation}
\label{PiV}
\begin{split}
P_1(V)& = \frac{1}{2} + \frac{1}{3}\,V\\
P_2(V)& = \frac{17}{18} + \frac{34}{27}\,V+ \frac{10}{27}\,V^2\\
P_3(V)& = \frac{457}{162} + \frac{457}{81}\,V+ \frac{266}{81}\,V^2+\frac{140}{243}\,V^3
\end{split}
\end{equation}

\subsection{Monoatomic gas}

In the case when  the particle-atom interaction has a power law tail \eqref{Ur} in the small separation limit, the joint distribution approaches a scaling form 
\begin{equation}
\label{scaling-fd-mono}
f(r,v,t)=(\Omega_d)^{-2} \left(\tau^\Lambda \sqrt{T}\,t\right)^{-d}\, F(V, R)
\end{equation}
with scaled spatial and velocity variables
\begin{equation}
\label{scaling-RV-mono}
R = \frac{r}{\sqrt{T}\,t}\,, \quad V = \frac{v}{\tau^\Lambda}
\end{equation}
The analog of equation \eqref{FRV} reads 
\begin{eqnarray}
\label{FRV-mono}
(1+\Lambda)dF\!\!&+\!\!&RF_R+\Lambda VF_V
+ V^{-\gamma}[(d-\gamma)F_V+VF_{VV}]\nonumber\\
 &+\!\!&\mathcal{D}\,\frac{V}{d}\!\left(\frac{d-1}{R}\,F_R+F_{RR}\right)=0
\end{eqnarray}
Here $\mathcal{D}$ is a numerical factor which quantifies diffusion in the Lorentz gas where the particle-scatters interaction is given by \eqref{Ur}.

Multiplying equation \eqref{FRV-mono} by $R^{i+d-1}V^{j+d-1}$ and integrating we arrive at the moment relations 
\begin{eqnarray}
\label{Mij-d-mono}
(i+\Lambda j)M_{i,j} &=& j(j+d-1-\gamma) M_{i,j-1-\gamma}\nonumber \\
&+&\mathcal{D}\,\frac{i(i+d-2)}{d}\,M_{i-2,j+1}
\end{eqnarray}
To the best of our knowledge, the value of the numerical constant $\mathcal{D}$ is not known. 

\section{Joint Distribution}
\label{full_solution} 

Here we derive the announced results \eqref{joint_d}--\eqref{density_d} by employing an approach based on the combination of the Laplace and Fourier transforms. It proves easier to deal with original kinetic equations \eqref{hydrodynamic-d} rather than with its scaled version. As a bi-product, we can also see that the solution approaches the scaling form. 

We begin again with the one-dimensional setting and show that the Laplace and Fourier transforms allow one to solve Eq.~\eqref{hydrodynamic-1d} for an arbitrary initial velocity distribution. Then we generalize to higher dimensions.

\subsection{One Dimension}

It is convenient to study Eq.~\eqref{hydrodynamic-1d} on the entire line $-\infty<x<\infty$ while for the velocity will be taken positive, $0\leq v<\infty$, as previously. Performing the Laplace transform in the $v$ variable and the Fourier transform in the $x$ variable, we find that the transformed joint distribution
\begin{equation}
\label{LF_def}
g(q, k, \tau)= \int_{-\infty}^\infty dx\,e^{iqx}\int_0^\infty dv\,e^{-vk}\,f(x,v,\tau)
\end{equation}
satisfies 
\begin{equation}
\label{LF_kinetic}
\frac{\partial g}{\partial\tau} + \left(k^2-Q^2\right)\frac{\partial g}{\partial k} = -k\,g,
\quad Q^2\equiv \frac{q^2}{4\rho^2 T}
\end{equation}
This linear hyperbolic partial differential equation can be solved using the method of characteristics. The characteristics are the curves in the $(k,\tau)$ plane which are found from 
\begin{equation}
\label{char_eq}
\frac{d k}{d \tau} = k^2 - Q^2
\end{equation}
Solving this differential equation we get
\begin{equation}
\label{char}
k = -Q\coth[Q(\xi + \tau)]
\end{equation}
where $\xi$ parameterizes different characteristics. Along a characteristics, that is keeping $\xi$ fixed, the governing equation \eqref{LF_kinetic} becomes
\begin{equation}
\label{g_char}
\frac{d g}{d \tau}\Big|_{\xi=\text{const}}= -k\,g
\end{equation}
Using \eqref{char} we express $k$ via $\xi$ and $\tau$, so that Eq.~\eqref{g_char} becomes
\begin{equation}
\label{g_charac}
\frac{d g}{d \tau} = Q\coth[Q(\xi + \tau)] g
\end{equation}
whose solution reads
\begin{equation}
\label{g_sol}
g = \sinh[Q(\xi + \tau)]\, G(\xi)
\end{equation}
Specializing \eqref{char} and \eqref{g_sol} to $\tau=0$ we get
\begin{equation*}
g_0(k,Q) = \sinh(Q\xi)\, G(\xi), \quad k= - Q\coth(Q\xi)
\end{equation*}
so that 
\begin{equation}
\label{G_sol}
G(\xi) = \frac{g_0[- Q\coth(Q\xi),Q]}{\sinh(Q\xi)}
\end{equation}
Combining \eqref{g_sol}--\eqref{G_sol}  we arrive at the exact solution for the transformed joint distribution 
\begin{equation}
\label{g_solution}
g = \frac{\sinh[Q(\xi + \tau)]}{\sinh(Q\xi)}\,g_0[- Q\coth(Q\xi),Q]
\end{equation}
Using \eqref{char}, we massage the ratio and rewrite the argument of $g_0$ to to transform \eqref{g_solution} into
\begin{equation}
\label{g_solution2}
g(k,q,\tau) = \frac{1}{\cosh s+\frac{k}{Q}\,\sinh s}\,g_0\left(\frac{k+Q\tanh(s)}{1+\frac{k}{Q}\tanh(s)},Q\right)
\end{equation}
where we have used the notation $s=Q\tau$ which has been used previously, e.g. in \eqref{cosh1}. 
This exact solution is valid for any initial distribution
\begin{equation}
g_0(k,q)= \int_{-\infty}^\infty dx\,e^{iqx}\int_0^\infty dv\,e^{-vk}\,f(x,v,\tau=0)
\end{equation}

Consider now the simplest initial velocity distribution
\begin{equation}
\label{delta_delta_1d}
f(x,v,\tau=0)=\delta(x)\,\delta(v)
\end{equation}
which corresponds to the initially stationary particle at the origin. The governing equation Eq.~\eqref{hydrodynamic-1d} is formally applicable if $v\gg \sqrt{T}$ (since the simplification of the collision integral in Eq.~\eqref{B-1d} leading to Eq.\eqref{hydrodynamic-1d} is valid only under this condition), but we are now more concerned with finding the simplest solution, in addition the initial condition is asymptotically irrelevant. For the initial condition \eqref{delta_delta_1d} we get $g_0=1$ and the transformed joint distribution becomes 
\begin{equation}
\label{g_deltas}
g =\frac{1}{\cosh s +\frac{k}{Q}\,\sinh s}\,, \quad s=Q\tau = q\sqrt{T}\,t
\end{equation}

The dependence on $k$ in \eqref{g_deltas} is very simple, so we perform the inverse Laplace transform and obtain
\begin{equation*}
f(x,v,\tau) = \int_{-\infty}^\infty \frac{dq}{2 \pi}\,e^{-iqx}\,\frac{s \, e^{-Vs \coth s} }{\tau \sinh s}
\end{equation*}

Note that the above formula already has the scaling form (for the initial condition \eqref{delta_delta_1d} the scaling form establishes instantaneously). Extending the variable $V$ to the whole axis (this amounts to replace $V\rightarrow|V|$ and divide by $2$) and re-writing the distribution in the manifestly scaling form ($f(x,v,\tau) = \frac{F(X,V)}{4 v_* x_*}$, see \eqref{scaling-f}) we get 
\begin{equation}
\label{joint_1d}
F(X,V) = \int_{-\infty}^\infty \frac{ds}{\pi} \, e^{-isX} \frac{s e^{-Vs\coth s}}{\sinh s}
\end{equation}
Integrating in velocity, $N(X)=\int_0^\infty dV\,F(X,V)$, we arrive at the announced result \eqref{NX}. 

\begin{figure}
\begin{tabular}{cc}
\includegraphics[width=0.9\columnwidth]{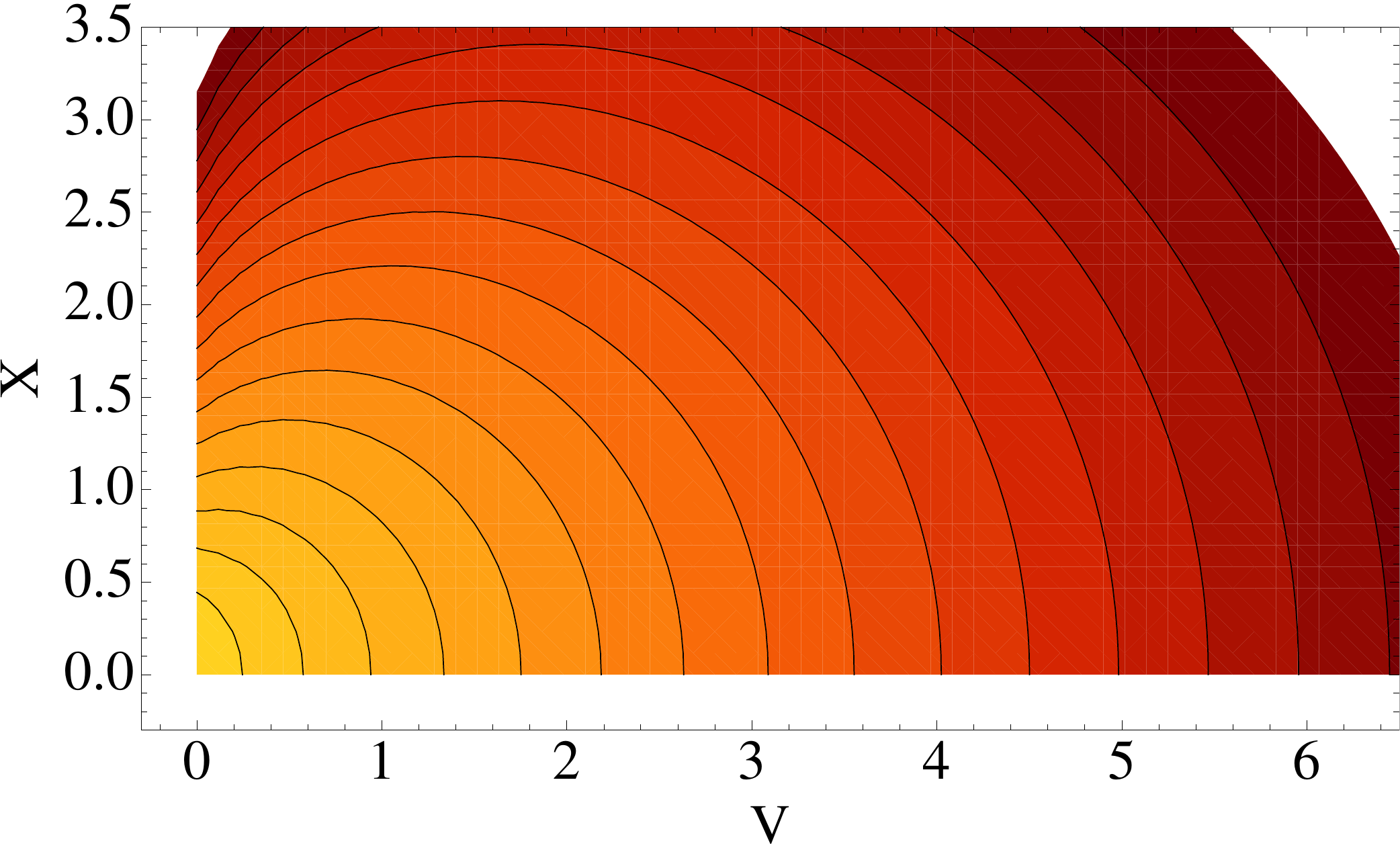} \\
\includegraphics[width=0.9\columnwidth]{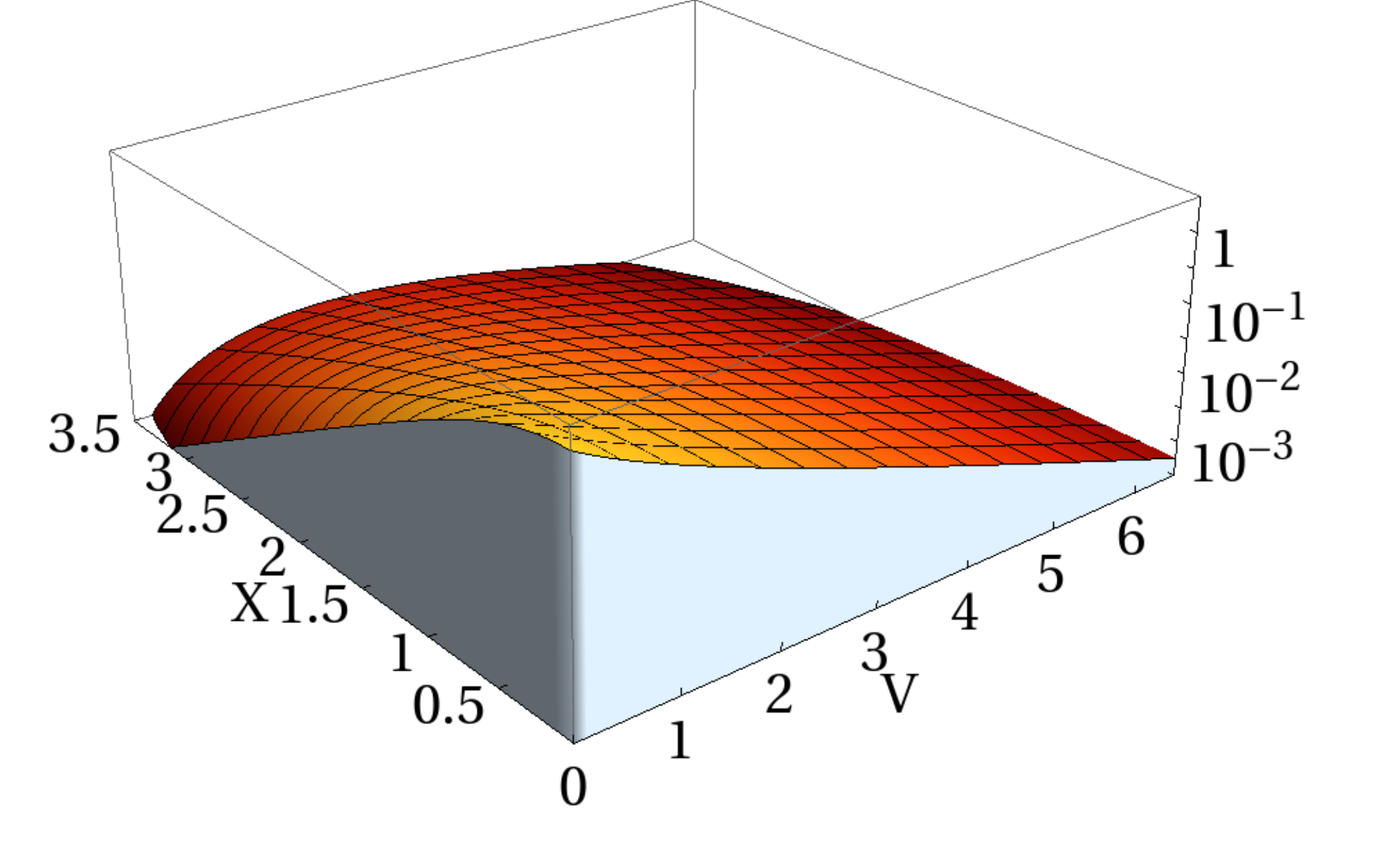}
\end{tabular}
\caption{Contour-plot (top panel) and 3D-plot (bottom panel) of the joint distribution function $F(X,V)$ for hard spheres gas in $1d$, Eq.~\eqref{joint_1d}.}
\label{joint-theory}
\end{figure}

\begin{figure}
\includegraphics[width=0.9\columnwidth]{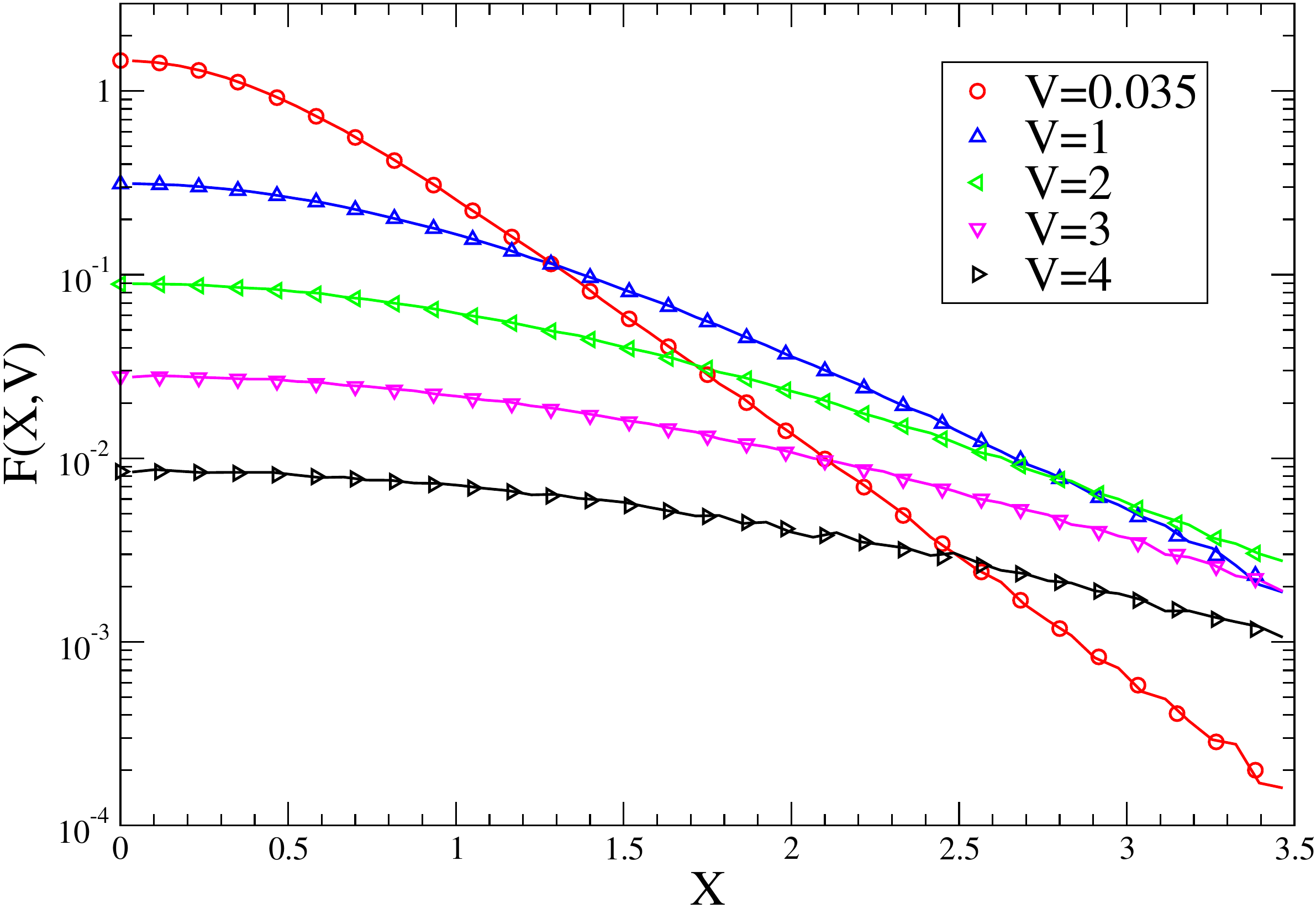} 
\caption{Values of $F(X,V)$ for hard spheres gas in $1d$ along the lines of fixed $V=0.0035,1,2,3,4$. The continuous lines are obtained from the numerical simulations (see Sec.~\ref{numerics}) while the symbols represent the values obtained by computing the integral \eqref{joint_1d}.}
\label{cutfixedV}
\end{figure}

\begin{figure}
\includegraphics[width=0.9\columnwidth]{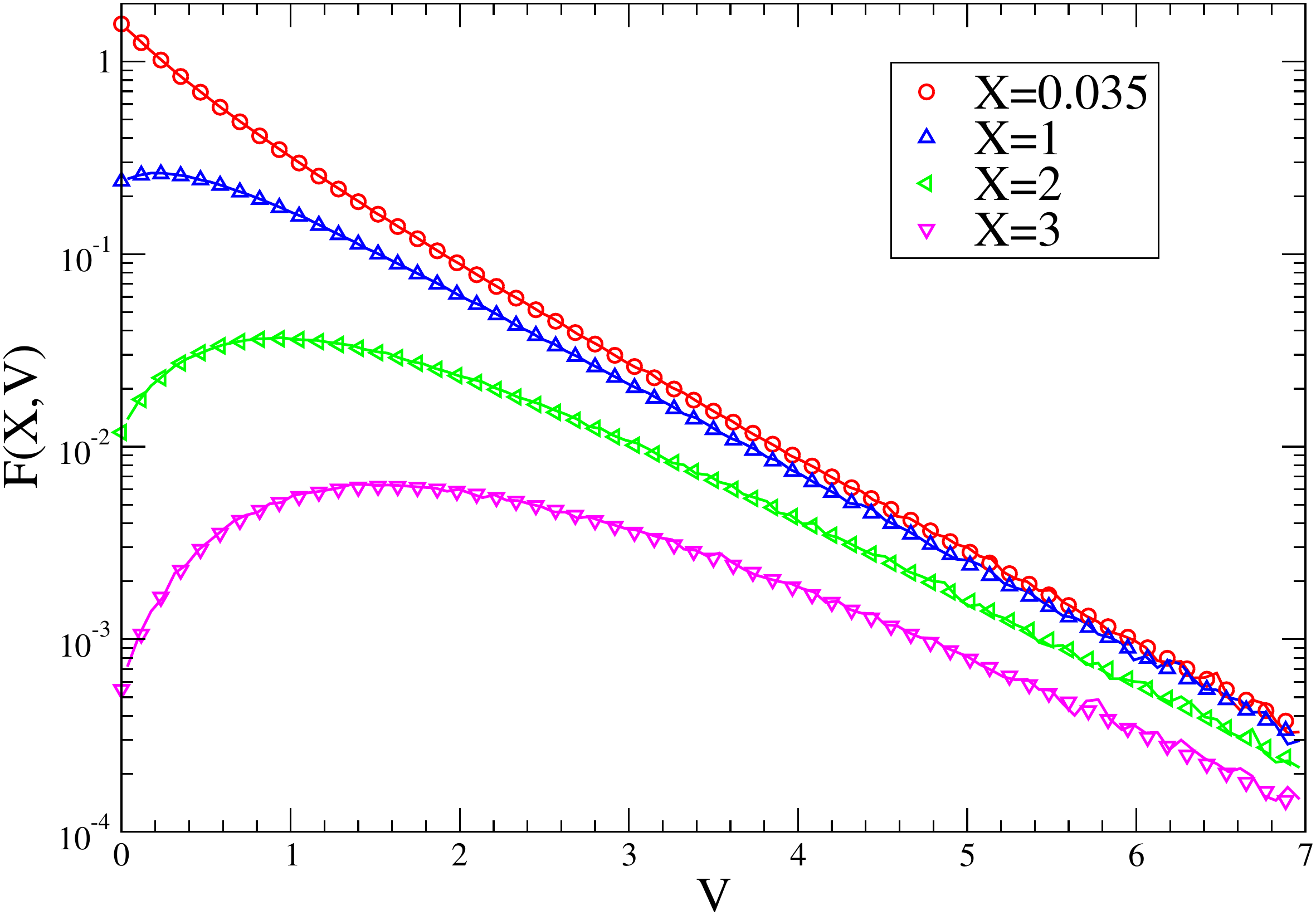} 
\caption{Values of $F(X,V)$ for hard spheres gas in $1d$ along the lines of fixed $X=0.0035,1,2,3$. 
The data are obtained as explained in Fig.~\ref{cutfixedV}.}
\label{cutfixedX}
\end{figure}

We could not compute the integral \eqref{joint_1d} in a closed form, so we determined it numerically. The results of the numerical integration (Fig.~\ref{joint-theory}) are in excellent agreement with the results of direct simulations (Fig.~\ref{joint}). 
The excellent agreement between theory and simulations is further shown in Fig.~\ref{cutfixedV}--\ref{cutfixedX} 
and provides a verification of our simulation scheme 
and shows that the replacement of the convection term by effective diffusion is indeed asymptotically exact.

\subsection{Higher Dimensions}

The joint distribution $f({\bf r}, {\bf v}, \tau)$ is isotropic in ${\bf r}$ and ${\bf v}$. It is convenient to explicitly assume the latter, so we want to find $f({\bf r}, v, \tau)$. We define the Laplace-Fourier transform of this distribution through 
\begin{equation}
\label{LF:general}
g({\bf q}, k,\tau)=\Omega_d\int d{\bf r}\,e^{i {\bf q}\cdot {\bf r}} \int_0^\infty dv\,v^{d-1}\,e^{-vk}\,f({\bf r}, v, \tau)
\end{equation}
We limit ourselves to the hard-sphere interaction. Applying the Laplace-Fourier transform to \eqref{hydrodynamic-d} we obtain
\begin{equation}
\label{LF_kinetic:d}
\frac{\partial g}{\partial\tau} + \left(k^2-Q^2\right)\frac{\partial g}{\partial k} = -dk\,g
\end{equation}
where we have used the short-hand notation
\begin{equation*}
Q^2 = \frac{q^2}{d(2Aa^{d-1}\rho)^2 T}\equiv \frac{{\bf q}\cdot {\bf q}}{d(2Aa^{d-1}\rho)^2 T}
\end{equation*}
The characteristics curves in the $(k,\tau)$ plane are defined by the same equation \eqref{char_eq} as in one dimension, while instead of \eqref{g_char}--\eqref{g_charac} we get
\begin{equation*}
\frac{d g}{d \tau}\Big|_{\xi=\text{const}}= -dk\,g = dQ\coth[Q(\xi + \tau)] g
\end{equation*}
Integrating we find
\begin{equation*}
g = \left(\sinh[Q(\xi + \tau)]\right)^d G(\xi)
\end{equation*}
while the general solution 
\begin{equation*}
\label{general_solution}
g(k,q,\tau) = \left(\cosh s+\frac{k}{Q}\,\sinh s\right)^{-d}\,g_0\left(\frac{k+Q\tanh(s)}{1+\frac{k}{Q}\tanh(s)}, Q\right)
\end{equation*}
with $s=Q\tau$. For the simplest initial velocity distribution
\begin{equation}
\label{delta_delta}
f({\bf r}, {\bf v},\tau=0)=\delta({\bf r})\,\delta({\bf v})
\end{equation}
the general solution simplifies to
\begin{equation}
\label{delta_sol}
g = \left(\cosh s+\frac{k}{Q}\,\sinh s\right)^{-d}
\end{equation}
As a check of this result we set $q=0$. Then $s=Q\tau=0$ and $\lim_{Q\to 0}Q^{-1}\sinh s=\tau$, so that Eq.~\eqref{delta_sol} becomes $g(k,q=0,\tau)=(1+\tau k)^{-d}$ which is exactly 
the Laplace transform of the velocity distribution [see \eqref{solution}]. 

Thus the joint distribution is the inverse Laplace-Fourier transform of \eqref{delta_sol}.  
Performing the inverse Laplace transform  of \eqref{delta_sol} in $k$ is easy. Therefore the final answer is the inverse Fourier transform. Re-writing the result in the scaling form we arrive at the announced scaled joint distribution \eqref{joint_d}. Similarly we obtain \eqref{density_d}. 

Equations \eqref{joint_d} and \eqref{density_d} involve integrals of the kind
\begin{equation}
\label{int:JR}
J({\bf R}) = \int d{\bf s}\, 
e^{-i \sqrt{d}\, {\bf s}\cdot {\bf R}}\, \Phi(s)
\end{equation}
The integral $J({\bf R})$ is actually rotationally invariant, $J({\bf R}) = J(R)$, which becomes clear by noting that we can simultaneously rotate ${\bf R}$ and ${\bf s}$. Using spherical coordinates we write
$d{\bf s} = \Omega_{d-1} (\sin\theta)^{d-2}s^{d-1}ds\,d\theta$ where $\theta$ is the angle between ${\bf s}$ and ${\bf R}$ (that is, we have ${\bf s}\cdot {\bf R} = sR\cos\theta$). This allows us to reduce the $d-$fold integral \eqref{int:JR} to the double-fold integral
\begin{equation*}
J(R) = \Omega_{d-1}\int_0^\infty ds\,s^{d-1}\,\Phi(s)\int_0^\pi d\theta\,(\sin\theta)^{d-2}\,e^{-i\sqrt{d}\,sR\cos\theta}
\end{equation*}
The integral in $\theta$ is computable, so one actually reduces \eqref{int:JR} to a single integral. 

For example in two dimensions we have
\begin{equation}
\label{joint_2}
F(R,V) = 2 \int_0^\infty ds\,s^3\,\frac{J_0(\sqrt{2}sR)}{(\sinh s)^2}\,e^{-Vs\coth s}
\end{equation}
and
\begin{equation}
N(R) = 2 \int_0^\infty ds\,s\,\frac{J_0(\sqrt{2}sR)}{(\cosh s)^2}
\label{NR2d}
\end{equation}
while in three dimensions we obtain
\begin{equation}
\label{joint_3}
F(R,V) = \frac{3}{\pi R} \int_0^\infty ds\,s^4\,\frac{\sin(\sqrt{3}sR)}{(\sinh s)^3} \,e^{-Vs\coth s} 
\end{equation}
and
\begin{equation}
\label{NR_3d}
N(R) = \frac{6}{\pi R} \int_0^\infty ds\,s\,\frac{\sin(\sqrt{3}sR)}{(\cosh s)^3} 
\end{equation}
Computing the integral on the right-hand side of \eqref{NR_3d} we arrive at the announced result \eqref{NR3}. The integrals defining the joint distribution in $d=2,3$ (Eq.\eqref{joint_2}--\eqref{joint_3}) were evaluated numerically and the resulting distributions are qualitatively similar to the one shown in Fig.~\ref{joint-theory}  for the $1d$ case.

\section{Numerical Simulations}
\label{numerics}

In order to verify our theoretical results we have used different types of numerical simulations. 

The most straightforward numerical approach to check our theoretical results would be to perform a full molecular dynamics (MD) simulation. We are interested, however, in the evolution of a single particle in a gas of background atoms. The MD simulations are very inefficient to study such a situation since they keep track and update the positions and velocities 
of {\it all} the background atoms that are unnecessary to compute the quantities of interest. 
Whenever possible we turn to less costly computational method.   

For the hard sphere gas in one and two dimensions, the {\it in-homogeneous}  Boltzmann equation was simulated by stochastically updating the velocity and positions of $10^6$ and $10^8$ particles respectively. A particle with velocity ${\bf v}$ travels for a time $\Delta t$ from the last collision covering a distance ${\bf v}\Delta t$ before colliding with a background atom with velocity ${\bf u}$. At the instant of collision the particle's velocity changes.
Thus the update rules are: 
\begin{subequations}
\begin{align}
t_{n+1}&=t_n + \Delta t_n 
\label{update1}\\
{\bf r}_{n+1}&= {\bf r}_n +{\bf v}_n \Delta t_n 
\label{update2}\\
{\bf v}_{n+1}&={\bf v}_n+2{\bf e}[({\bf u}-{\bf v}_n)\cdot{\bf e}] 
\label{update3}
\end{align}
\end{subequations}

Under the assumption already used in writing down the Lorentz-Boltzmann equation, the quantities 
$\Delta t, {\bf u}, {\bf e}$ are random variables whose distributions need to be specified in order to
have a complete description of the temporal evolution. The velocity update rule \eqref{update3} can be understood by analyzing the collisions in the reference frame of the background atom (which in our case coincides with the center of mass reference frame).  The key feature of the hard-sphere interaction is that the collision rate is proportional to the absolute value of the relative velocity $g$, so 
that the particle more often collides with atoms moving in direction opposite to its own. 

The random variable $\Delta t$ is the first collision time which is distributed according to a Poisson  process. This can be understood in the following way. The particle can collide with any background atom. 
The probability that the particle has not collided with the background atom $i^{\rm th}$ up to time $t$ is called $S_i(t)$. The survival probability $S_i(t)$ is decaying in time and satisfies a very simple differential equation:
\begin{equation}
\label{survival}
\frac{\partial S_i(t)}{\partial t}=- r_i \, S_i(t), \quad r_i \sim |{\bf v}-{\bf u}_i|
\end{equation} 
The rate of collision, $r_i$, is proportional to the absolute value of the relative velocity with respect the 
$i^{\rm th}$ atom. The probability that the particle has not collided with {\it any} atom up to time $t$ is $S(t)=\prod_{i=1}^{N} S_i(t)$, where $N$ is the total number of background atoms.
Using Eq.~\eqref{survival} and the definition of $S(t)$ we obtain
\begin{equation}
\label{survival-tot}
\frac{\partial S(t)}{\partial t}=- r \,  S(t), \quad r=\sum_{i=1}^{N} r_i
\end{equation} 
whose solution is a simple exponential decay with rate $r$. Note that $S(t)$ is also the probability that the first collision happens at time $t$, i.e. $S(t)$ is the distribution of the first collision time. Reintroducing the dependence on the particle velocity explicitly we obtain the probability $P(\Delta t|{\bf v})$ that the particle with velocity ${\bf v}$ collides for the first time at time $\Delta t$:
\begin{subequations}
\begin{align}
P(\Delta t|{\bf v})&= r({\bf v})\,\exp(-r({\bf v}) \Delta t) 
\label{Dt-1}\\
r({\bf v}) & =\sum_{i=1}^{N} r_i({\bf v})= 2 a \rho \,\langle |{\bf v}-{\bf u}|\rangle_{\bf u}
\label{Dt-2}
\end{align}
\end{subequations}
Here $\langle (\cdot)\rangle_{\bf u}$ denotes the average over the velocity distribution of the background atoms, $a$ is the radius of the hard-spheres and $\rho$ is the number density of background atoms. 
The last equality in \eqref{Dt-2} has been specified for the two-dimensional case. 

The probability of making the first collision with the $i^{\rm th}$ atom is \cite{decay} 
\begin{equation}
\label{ratio}
\frac{r_i({\bf v})}{r({\bf v})}=\frac{|{\bf v}-{\bf u}_i|}{N \langle |{\bf v}-{\bf u}|\rangle_{\bf u}}
\end{equation}
This equation can be understood in the following way. 
If it was equally likely to collide with any atom only the factor $1/N$ would appear in Eq.~\eqref{ratio}.
The correction ($\frac{|{\bf v}-{\bf u}_i|}{\langle |{\bf v}-{\bf u}|\rangle_{\bf u}}$) 
in Eq.~\eqref{ratio} to this simple behavior describes the fact the the particle collides preferentially with atoms moving in direction opposite to its own.
It is worth noting that this correction approaches $1$ if $v\gg \langle u \rangle$.

The calculation of the total rate is difficult in any dimension $d>1$. It can be approximated by
\begin{equation}
\label{rtot_simple}
\langle |{\bf v}-{\bf u}|\rangle_{\bf u} \sim |{\bf v}|+\langle|{\bf u}|\rangle_{\bf u} \sim |{\bf v}|+\sqrt{T}
\end{equation}
Only the limiting behavior for $v\gg\sqrt{T}$ and $v\ll\sqrt{T}$ of Eq.~\eqref{rtot_simple} are important.
We are interested in the large time limit when $v\gg \langle u \rangle$ and $r({\bf v})\sim N|{\bf v}|$. Equation \eqref{rtot_simple} correctly reproduces this limit. Moreover, Eq.~\eqref{rtot_simple} ensures that a particle with an unexpected low velocity (in the extreme case ${\bf v}=0$) will collide with a 
background atom with a rate proportional to the thermal velocity of the background gas.
 
Using \eqref{Dt-1}--\eqref{rtot_simple} one computes the collision time $\Delta t$. 
Then a background velocity ${\bf u}$ is generated from the Maxwell-Boltzmann distribution \eqref{MB-1d} 
and it is accepted with probability $\frac{|{\bf v}-{\bf u}|}{\langle |{\bf v}-{\bf u}|\rangle_{\bf u}}$ (see Eq.~\eqref{ratio}). Finally the random variable ${\bf e}$ is generated from the distribution \eqref{De}.

\begin{figure}
\includegraphics[width=0.9\columnwidth]{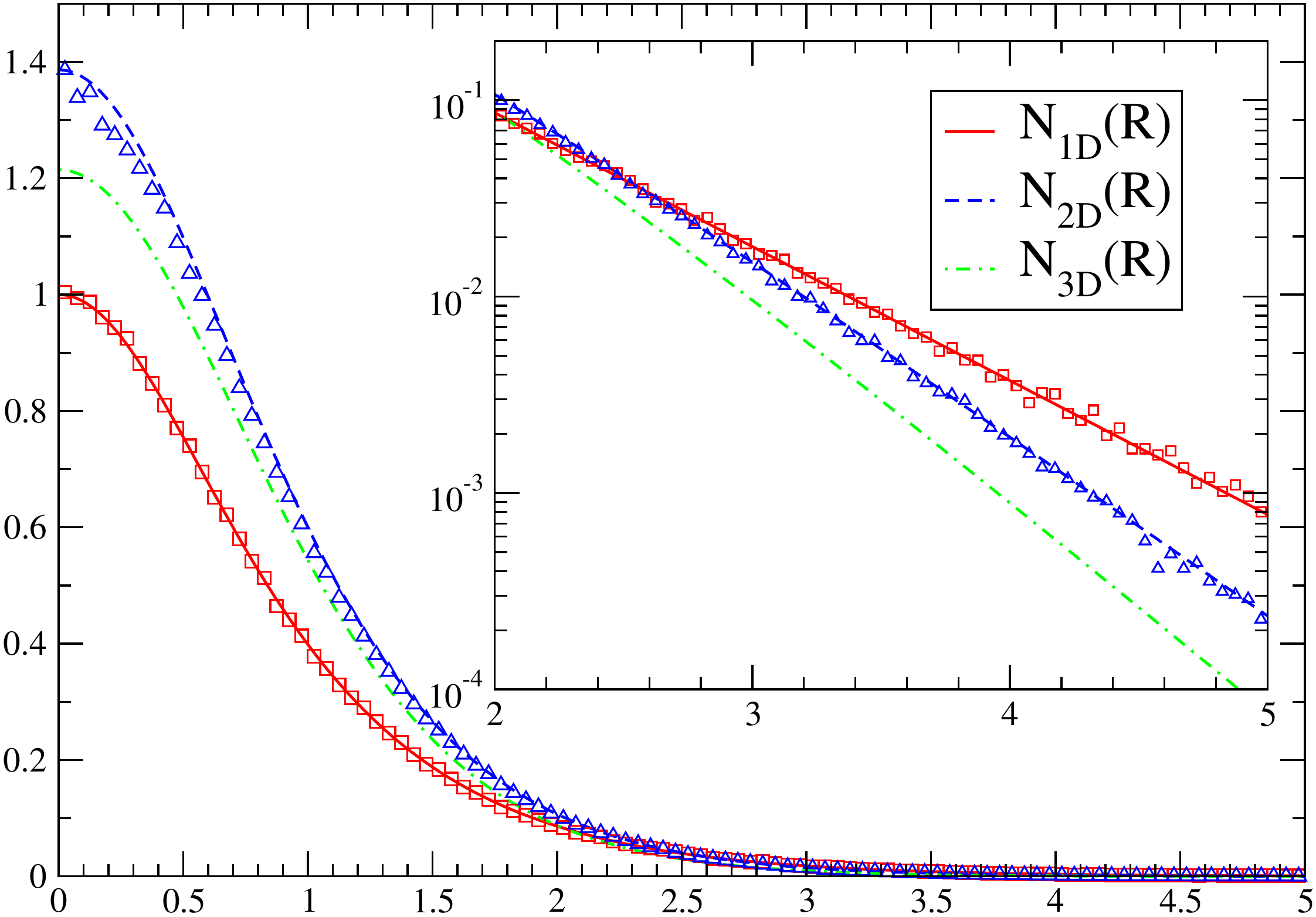}
\caption{Density profile for the hard sphere gas vs. the rescaled variable
$R=r/\sqrt{T} t$. 
The numerical simulations in $d=1,2$ are compared with the theoretical predictions, Eq.~\eqref{NX} and Eq.~\eqref{NR2d} (integrated numerically).
The theoretical prediction for $d=3$, Eq.~\eqref{NR3}, is also shown.}
\label{r}
\end{figure}

The velocity distribution is in excellent agreement with the exponential scaling form. 
The density profiles are shown in Fig.~\ref{r}. In one dimension, there is a perfect agreement with the theoretical prediction, Eq.~\eqref{NX}. In two dimensions, the numerical simulation correctly reproduces the known values for the moments $\langle R^{2n}\rangle$ (see Table~\ref{momentstable}) 
and agrees with the prediction \eqref{Nd-exp-tail} for the tail. 

In the one-dimensional case, every velocity distribution of the background atoms is stationary (since in a two-body collision the atoms merely exchange their velocities). In particular it is possible to chose a uniform velocity distribution for $-u_{max}< u < u_{max}$. In this case the total rate (Eq.~\eqref{Dt-2}) can be calculated exactly and Eq.~\eqref{ratio} can be enforced very efficiently. In this situation we were able to stochastically update the velocity and positions of $10^8$ particles which allowed us to simulate the joint distribution $F(X,V)$ (see Fig.~\ref{joint}). It is interesting to note how the exponential character of the speed distribution $F(V)$ is also present for $F(V,X=0)$. 
In the same way the character of the density distribution $N(X)$ persists for $F(V=0,X)$. 
In the contour-plot (top panel of Fig.~\ref{joint}) we observe that the equiprobability line always cross the $V$-axis perpendicularly while they cross the $X$-axis at acute (obtuse) angle for $X<X_c$ ($X>X_c)$ where $X_c \sim 0.8$. This has the consequence that for any given velocity
the maximum probability is always at $X=0$ (green line in Fig.~\ref{joint}) 
while for fixed $X$ the maximum probability is at $V=0$ only for $X<X_c$ (blue line in Fig.~\ref{joint}). 
The numerical result clearly show the lack of factorization: The joint distribution $F(X,V)$ is not a product of  functions of $X$ and $V$.    

In two dimensions, we have also used a ``brute-force'' molecular dynamics simulations to investigate 
the case when the atoms interact between themselves and with the particle through the potential $U\sim r^{-\lambda}$. This simulation schemes is much more time-consuming than the stochastic update of the position and velocity of the particle. 
For this reason we were able to simulate only $10^4$ particles. This is sufficient to check the scaling of the average velocity and displacement with time, but does not allow us to check the full distribution. 
In our system the background atoms are affected by other atoms and insensitive to the presence of the particle; the particle is affected by the atoms. Computationally this property is implemented in a simple way. At each time step of the molecular dynamic simulation we calculate the total force acting on a background atom summing {\it only} the contributions 
from the other atoms (no contribution from the particle). 
The total force acting on the particle is obtained summing {\it all} the contributions from the atoms. 

Numerically it is convenient to simulate many independent particles in the same background gas of atoms. Usually, even if the particle-particle potential is set to zero, particles interact indirectly via the background gas. In our case, the particles do not affect the background atoms and are totally independent from each other. We have simulated $10^4$ independent particles in the same background gas of $5\cdot 10^3$ atoms. For the reason explained before this simulation scheme is equivalent to $10^4$ runs of a single particle in a background gas of $5\cdot 10^3$ atoms.

The equations of motion have been numerically integrated using the velocity-Verlet algorithm \cite{verlet}. 
The time-step of the numerical integration was reduced during the time evolution in order to keep the average particle's displacement during a single time step constant and smaller than the mean free-path of the gas. The initial positions of the background atoms and of the particles were randomly drawn from the uniform distribution inside the simulation box with periodic boundary conditions.
The initial velocity of the particles were drawn from the distribution $\delta (v-v_0)/{2 \pi}$ while 
the initial velocity of the atoms were generated from the Maxwell-Boltzmann distribution and were rescaled in order to ensure that the total energy ($\sim T$) of the background gas had a fixed value.

\begin{figure}
\includegraphics[width=0.9\columnwidth]{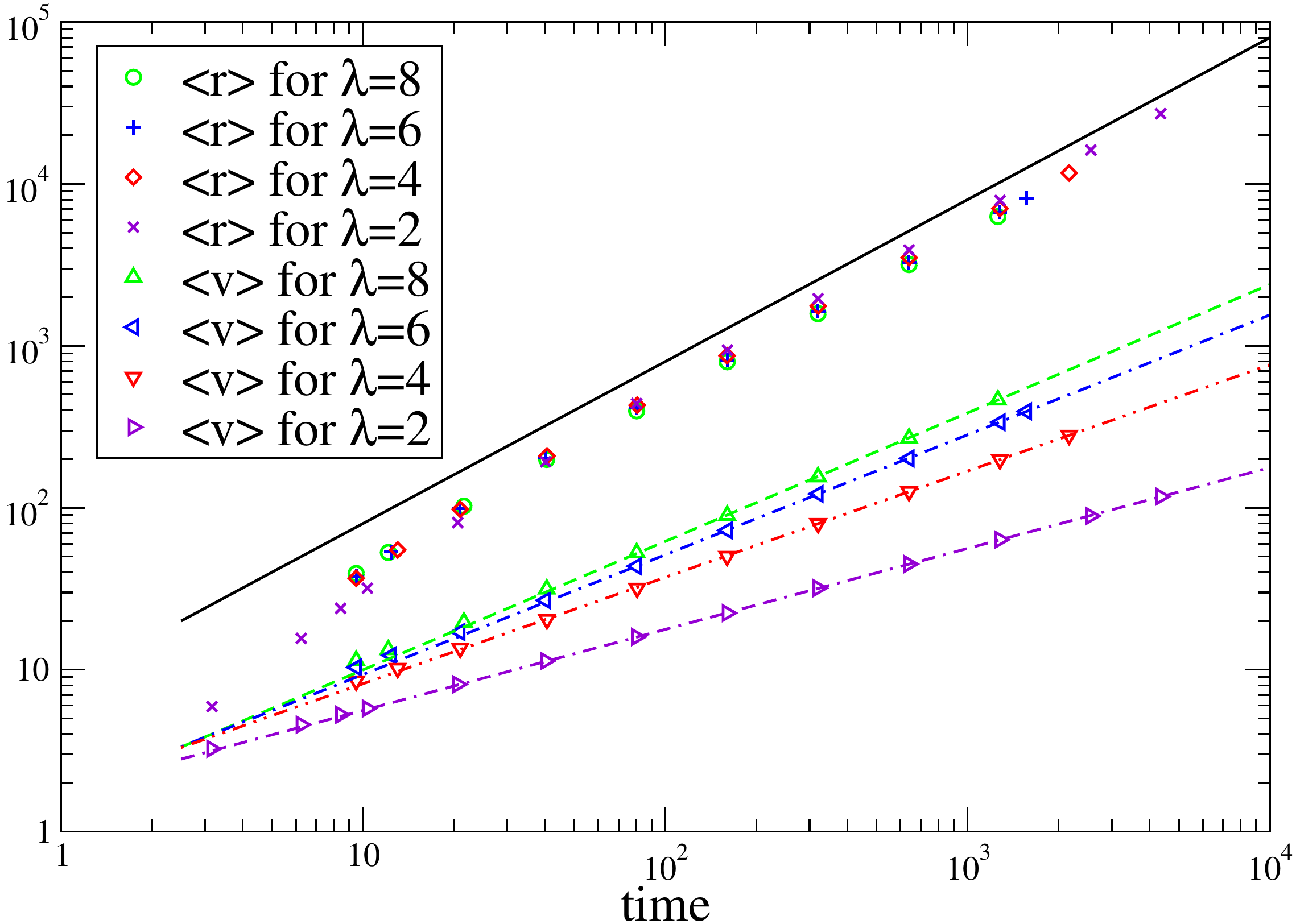}
\caption{Average particle velocity and displacement in two dimensions 
with a particle-atom interaction potential diverging as $U\simeq r^{-\lambda}$. 
In all cases the density of the background gas is $\rho =25\%$. 
The slopes of the fitting curves (dashed lines) are $0.5, 0.66, 0.74, 0.79$ (bottom to top), 
all in excellent agreement with the theoretical prediction $\lambda/(\lambda+2)$.
The solid line has slope $1$ and it is a guide for the eye.}
\label{samerho}
\end{figure}

\begin{figure}
\includegraphics[width=0.9\columnwidth]{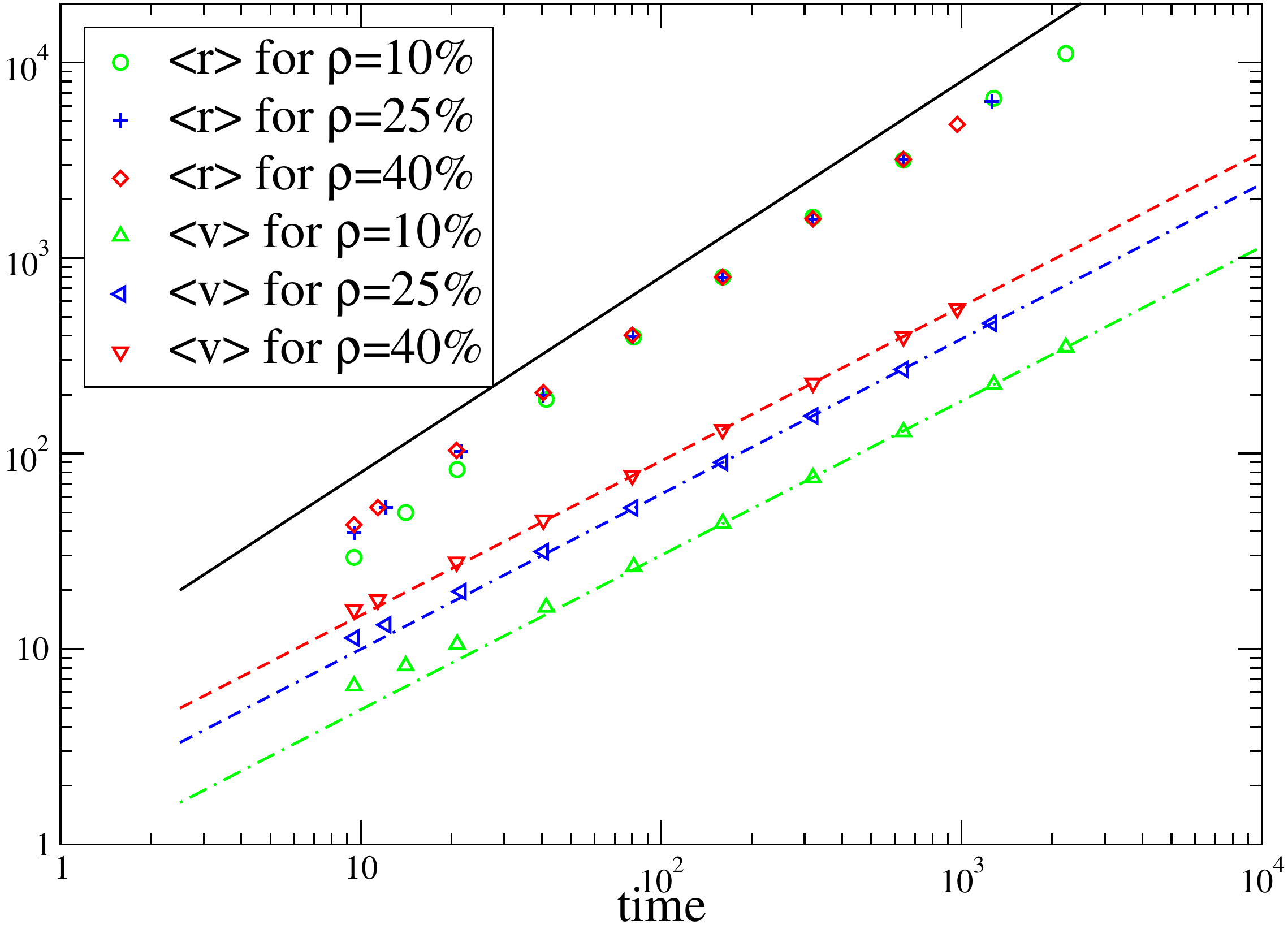}
\caption{The same system as in Fig.~\eqref{samerho} with fixed  interaction exponent 
$\lambda =8$ and varying density. 
The slope of the fitting curves (dashed lines) is $\Lambda=0.79$ in all cases,
while the intercepts are $b=0.79,1.60,2.41$ (bottom to top).
Note these values are in the ratio $1:2.02:3.05$ in excellent 
agreement with the theoretical prediction \eqref{final_v}
$b_i/b_j=(\rho_i/\rho_j)^\Lambda$ which gives $1:2.08:3.03$.
The solid line has slope $1$ and it is a guide for the eye.}
\label{samelambda}
\end{figure}

The results of different simulations at fixed density and fixed interaction exponent 
are shown in Fig.~\ref{samerho} and \ref{samelambda} respectively 
and are in excellent agreement with theoretical predictions.

Finally, the quasi-recurrent relation \eqref{Mij-d-mono} has been iteratively solved (as shown in Fig.~\ref{recurrence} and explained in the text) with {\it Mathematica} to calculate exactly the moments of the spatial distribution $\langle R^{2n}\rangle$ up to $n=500$ for the hard sphere gas in $d=1,2,3$. 

In Fig.~\ref{saturation} we show the ratio $(2n)^2 \langle R^{2n}\rangle / \langle R^{2(n+1)}\rangle$ which allows us to extract the asymptotic exponential decay of the density distribution.

\section{Summary}
\label{summ}

We have analyzed the behavior of a very light particle in an equilibrium background gas. We have shown that in the long-time limit, the average particle displacement grows linearly with time and proportionally to the thermal velocity of the background atoms --- the density of the gas, the size of atoms, and the details of the interaction between the particle and the atoms do not affect the asymptotic. The average particle velocity also grows in a rather universal way and the scaled velocity distribution approaches a scaling form which is generically non-Gaussian (the only exception is when the particle-atoms interaction is described by a Maxwell potential).  

For the hard-sphere particle-atom interaction in arbitrary dimensions, we have computed the asymptotically exact velocity distribution, position distribution and joint velocity-position distribution. The most complete results for the joint distribution have been derived using a combination of Fourier and Laplace transforms. 

In one dimension, we have also determined the probability density for the particle displacement using a less standard moment approach. Specifically, we have guessed an exact expression for the moments $\langle r^{2n}\rangle$, which we verified by exact ({\it Mathematica}--assisted) calculations of the moments up to $\langle r^{1000}\rangle$, and we found the probability density that results in these moments. We have also guessed an exact expression for the moments $\langle r^{2n}\rangle$ in two dimensions and we have confirmed to the same depth as in one dimension. Further, we have used the moments to establish the large displacement tail of the probability density and to study the correlations between the velocity and displacement of the particle. 

Our theoretical predictions are in perfect agreement with the numerical simulations providing strong evidence that our simulation scheme is correct and that the simplification of the collision integral and the replacement of the convective term by effective diffusion are indeed asymptotically exact in the limit when the particle velocity greatly exceeds the thermal velocity of atoms.

The Lorentz model was originally suggested \cite{LG} as an idealized model of electron transport. Quantum mechanics is of course essential for this problem.  In the context of the quantum particle in a container of fixed volume with boundaries deforming in a chaotic manner (a stochastic model for Fermi's acceleration of the quantum particle), some mostly numerical work has been done (see e.g. \cite{Doron}). Perhaps the most interesting extension of the present work is to analyze the quantum version of our model.

\vskip 0.2cm
We thank A. Polkovnikov for fruitful discussions. We
acknowledge support from NSF grant CCF-0829541 (PLK) and DOE grant DE-FG02-08ER46512 (LD'A).

\appendix
\section{Approach to Scaling}
\label{approach}

In one dimension, atoms merely exchange their velocities, so there is no relaxation and {\em any} velocity distribution $P(u)$ can be taken as an equilibrium distribution. As an example, consider the bimodal velocity distribution
\begin{equation}
\label{bimodal}
P(u) = \delta\!\left(u-\frac{1}{2}\right) + \delta\!\left(u+\frac{1}{2}\right)
\end{equation}
(The bimodal distribution is often used in studies of the one-dimensional Boltzmann equation, see e.g. \cite{PS}.) Note that for the bimodal velocity distribution the condition of Eq.~\eqref{Pu-1} holds; further, the density and the temperature of the background gas are $\rho=2, T=1/4$. Therefore $\tau=2\rho T t=t$ and the scaling solution \eqref{scaling-1d} becomes 
\begin{equation*}
f(v,t) = \frac{1}{2 t}\,e^{-|v|/t}
\end{equation*}

Let us now try to establish exact results starting with initial condition 
\begin{equation}
\label{in-delta}
f(v,t=0)=\delta(v)
\end{equation}
The velocity distribution cannot approach the smooth distribution \eqref{scaling-1d}. For the bimodal velocity distribution \eqref{bimodal} and the initial condition \eqref{in-delta}, the particle velocity can be only integer:
\begin{equation}
\label{integer}
f(v,t)=\sum_{n=-\infty}^\infty  P_n(t)\,\delta(v-n)
\end{equation}
The amplitudes $P_n(t)$ are still expected to behave as 
\begin{equation}
\label{scaling-Pn}
P_n(t) = \frac{1}{2 t}\,e^{-|n|/t}
\end{equation}
in the limit $|n|\to\infty$ and $ t\to\infty$, with $n/t$ being finite. 

To probe the exact behavior we insert \eqref{bimodal} and \eqref{integer} into the Boltzmann equation  \eqref{B-1d} and deduce an infinite set of rate equations
\begin{equation}
\label{Pn}
\dot P_n = \left(n-\frac{1}{2}\right)P_{n-1}+\left(n+\frac{1}{2}\right)P_{n+1}-2nP_n
\end{equation}
for $n\geq 1$ and 
\begin{equation}
\label{P0}
\dot P_0 = P_1 - P_0
\end{equation}
(It suffices to consider $P_n$ with $n\geq 0$; with initial condition \eqref{in-delta}, the particle velocity is a manifestly even function of $v$ and therefore 
$P_{-n}\equiv P_n$.)

To treat \eqref{Pn}-- \eqref{P0} let us use the generating function
\begin{equation}
\label{P-gen}
\mathcal{P}(t, z) = P_0(t)+ 2\sum_{n\geq 1} P_n(t)\,z^n 
\end{equation}
Note that 
\begin{equation}
\label{P-z1}
\mathcal{P}(t, z=1) = \sum_{n=-\infty}^\infty P_n(t) = 1
\end{equation}
explaining why we have chosen the definition \eqref{P-gen} of the generating function instead of $\sum_{n\geq 0} P_n(t)\,z^n$. 

Utilizing the generating function \eqref{P-gen} we recast an infinite set of rate equations \eqref{Pn}-- \eqref{P0} into a single partial differential equation
\begin{equation}
\label{long}
\frac{\partial \mathcal{P}}{\partial t} 
= (1-z)^2\,\frac{\partial \mathcal{P}}{\partial z} + \frac{(1-z)^2}{2z}\,P_0(t)
- \frac{1-z^2}{2z}\, \mathcal{P}
\end{equation}
We want to solve \eqref{long} subject to the initial condition $P_n(t=0)=\delta_{n,0}$, or equivalently 
\begin{equation}
\label{in-P}
\mathcal{P}(t=0, z)=1
\end{equation}
and the boundary condition \eqref{P-z1}. 

Using $\zeta=1/(1-z)$ instead of $z$, we re-write \eqref{long} as
\begin{equation}
\label{long-zeta}
\frac{\partial \mathcal{P}}{\partial t} 
= \frac{\partial \mathcal{P}}{\partial \zeta} + \frac{P_0(t)}{2\zeta(\zeta-1)}
+ \frac{1-2\zeta}{2\zeta(\zeta-1)}\, \mathcal{P}
\end{equation}
The transformation $\xi = (t+\zeta)/2,  \eta = (t-\zeta)/2$
recasts \eqref{long-zeta} into 
\begin{equation}
\label{long-eta}
\frac{\partial \mathcal{P}}{\partial \eta} 
=\frac{P_0(\xi+\eta) + [1-2(\xi-\eta)] \mathcal{P}}{2(\xi-\eta)(\xi-\eta-1)}
\end{equation}
To solve  \eqref{long-eta} we note that its homogeneous version, 
\begin{equation*}
\frac{\partial \mathcal{P}}{\partial \eta} 
=\frac{1-2(\xi-\eta)}{2(\xi-\eta)(\xi-\eta-1)}\, \mathcal{P}\,,
\end{equation*}
has a general solution 
\begin{equation*}
\mathcal{P}(\xi, \eta) = \sqrt{(\xi-\eta)(\xi-\eta-1)}\,Q(\xi)
\end{equation*}
where $Q(\xi)$ is an arbitrary function of $\xi$. Then a solution to the full equation \eqref{long-eta} can be sought using the variation of constant technique. In the present case we must actually vary the function $Q(\xi)$, namely, we should seek a solution of the form 
\begin{equation}
\label{PQ}
\mathcal{P}(\xi, \eta) = \sqrt{(\xi-\eta)(\xi-\eta-1)}\,Q(\xi,\eta)
\end{equation}
Plugging \eqref{PQ} into \eqref{long-eta} we obtain a simple equation for $Q$ which is integrated to find a final solution. Returning back to the variables $(t,\zeta)$ we get
\begin{equation}
\label{PQ-again}
\mathcal{P}(t, \zeta) = \sqrt{\zeta(\zeta-1)}\,Q(t,\zeta)
\end{equation}
with
\begin{equation}
\label{Q-sol}
\begin{split}
Q&=\frac{1}{ \sqrt{(t+\zeta)(t+\zeta-1)}}\\
&+ \frac{1}{2}\int_0^t d\tau\,\frac{P_0(\tau)}{[(t-\tau+\zeta)(t-\tau+\zeta-1)]^{3/2}}
\end{split}
\end{equation}

Equations \eqref{PQ-again}--\eqref{Q-sol} give rather formal results as we haven't yet extracted $P_0(t)$. However, on this stage we can already confirm the emergence of scaling \eqref{scaling-Pn}. Indeed, assuming that $P_0(t)$ decays and approaches to zero as $t\to\infty$, we conclude that the integral term on the right-hand side of \eqref{Q-sol} is asymptotically negligible and therefore 
$Q\simeq 1/(t+\zeta)$. Therefore \eqref{PQ-again} becomes $\mathcal{P}\simeq \zeta/(t+\zeta)$, where we additionally consider the large $\zeta$ limit. 
Hence $\mathcal{P}\simeq 1/(1+t/\zeta)=1/(1+t - tz)$. Expanding this result we get
\begin{equation*}
P_n(t) = \frac{1}{2}\frac{t^{n-1}}{(1+t)^n}
\end{equation*}
which in the scaling limit $n\to\infty$ and $ t\to\infty$ with $n/t$ being finite is indeed equivalent to \eqref{scaling-Pn}.

\section{Angular Integrals}
\label{app}

Let us first prove the validity of relation \eqref{int_A} with $A$ defined in \eqref{AB:def}. The integral in \eqref{int_A} is equal to $({\bf J}\cdot {\bf u})$, where ${\bf J} =  \int \mathcal{D}{\bf e}\,({\bf g}\cdot{\bf e})\, {\bf e}$. Due to symmetry, the vector ${\bf J}$ must be directed along ${\bf g}$. Hence
\begin{equation}
\label{Jg}
{\bf J} =  A{\bf g}
\end{equation}
where the amplitude $A$ is independent on $g$ since ${\bf J}$ scales linearly with $g$. Computing the scalar product of ${\bf g}$ and ${\bf J}$ we obtain
\begin{equation}
\label{A-gen}
A=\frac{1}{g^2}\,({\bf J}\cdot {\bf g}) =  
\frac{1}{g^2}\int \mathcal{D}{\bf e}\, ({\bf g}\cdot{\bf e})^2
\end{equation}
Using \eqref{Jg} we arrive at
\begin{equation*}
 \int \mathcal{D}{\bf e}\,({\bf u}\cdot{\bf e})({\bf g}\cdot{\bf e})
 = ({\bf u}\cdot {\bf J}) = A ({\bf u}\cdot {\bf g})
\end{equation*}
which together with \eqref{A-gen} lead to \eqref{int_A}. 

To establish \eqref{int_B} with $B$ defined in \eqref{AB:def}
we note that the integral in Eq.~\eqref{int_B} is equal to 
$({\bf u}\cdot \mathbb{T}\cdot {\bf u})$, where
\begin{equation}
\label{T-def}
\mathbb{T} =  \int \mathcal{D}{\bf e}\, ({\bf g}\cdot{\bf e})^2 {\bf e} {\bf e}
\end{equation}
Tensor $\mathbb{T}$ depends only on vector ${\bf g}$, so it must read
\begin{equation}
\label{Tg}
\mathbb{T} =  C_1{\bf g}{\bf g} + C_2 g^2\mathbb{U} 
\end{equation}
where $\mathbb{U}$ is the unit tensor. To determine the amplitudes $C_1$ and $C_2$, we compute the trace of tensor $\mathbb{T}$ and the product $({\bf g}\cdot \mathbb{T}\cdot {\bf g})$. Using \eqref{Tg} we find
\begin{subequations}
\label{ts1}
\begin{align}
{\rm Tr} (\mathbb{T}) &=  (C_1  + dC_2) g^2
\label{trace}\\
({\bf g}\cdot \mathbb{T}\cdot {\bf g}) &= (C_1  + C_2) g^4
\label{scalar}
\end{align}
\end{subequations}
If instead we use \eqref{T-def} we get
\begin{subequations}
\label{ts2}
\begin{align}
{\rm Tr} (\mathbb{T}) &= \int \mathcal{D}{\bf e}\, ({\bf g}\cdot{\bf e})^2 = A g^2
\label{trace2}\\
({\bf g}\cdot \mathbb{T}\cdot {\bf g}) &= 
\int \mathcal{D}{\bf e}\, ({\bf g}\cdot{\bf e})^4 = B g^4
\label{scalar2}
\end{align}
\end{subequations}
where we have used the definitions of $A$ and $B$, see \eqref{AB:def}. Comparing \eqref{ts1} with \eqref{ts2} we express the amplitudes $C_1$ and $C_2$ via $A$ and $B$:
\begin{equation}
C_1=\frac{dB-A}{d-1}\,,\quad C_2=\frac{A-B}{d-1}
\end{equation}
yielding indeed \eqref{int_B}. 

For the three-dimensional hard-sphere gas, the integration measure is given by Eq.~\eqref{De} and therefore 
\begin{equation}
\label{AB:HS}
\begin{split}
A &= \frac{1}{g^3} \int d^2{\bf e}\,\,\theta({\bf g}\cdot{\bf e})\,({\bf g}\cdot{\bf e})^3\\
B &=\frac{1}{g^5} \int d^2{\bf e}\,\,\theta({\bf g}\cdot{\bf e})\,({\bf g}\cdot{\bf e})^5
\end{split}
\end{equation}
Let us now introduce spherical coordinates with the axis along ${\bf g}$. We have
$d^2{\bf e}=2\pi\,\sin\vartheta\,d\vartheta$, $({\bf g}\cdot{\bf e})=g\,\cos\vartheta$; the term $\theta({\bf g}\cdot{\bf e})$ limits the integration over the range $0\leq \vartheta\leq \pi/2$. Thus
\begin{equation*}
A=2\pi \int_0^{\pi/2}\sin\vartheta\,(\cos\vartheta)^3\,d\vartheta
= \frac{\pi}{2}
\end{equation*}
and similarly $B=\pi/3$. Thus we obtain \eqref{integralA}--\eqref{integralB}.  (See Ref.~\cite{fluid} for the computation of integrals similar to \eqref{integrals}; such integrals often appear in kinetic theory of the hard-sphere gas.)

For the $d-$dimensional hard-sphere gas, we have the same expression \eqref{AB:HS} for $A$ and $B$, the only difference is that $d{\bf e}=\Omega_{d-1}\,(\sin\vartheta)^{d-2}\,d\vartheta$. Computing $A$ yields
\begin{equation}
\label{Ad}
A= \Omega_{d-1}\int_0^{\pi/2} (\sin\vartheta)^{d-2}\,(\cos\vartheta)^3\,d\vartheta
= \frac{\pi^{(d-1)/2}}{\Gamma(\frac{d+3}{2})}
\end{equation}

\section{Exact Solution of Eq.~\eqref{final_v} and Analysis of Solutions of Eq.~\eqref{kinetic}}
\label{exact}

Let us first solve Eq.~\eqref{final_v} using the Laplace transform. 
Note that in Eq.~\eqref{final_v} the variable $v$ varies in the range $(0,+\infty)$ and therefore we use the Laplace transform rather than e.g. the Fourier transform. 
In any number of dimension we define
\begin{equation}
\label{laplace_def}
g(k,\tau)=\Omega_d\,\int_0^\infty dv\,v^{d-1}\,e^{-vk}\,f(v,\tau)
\end{equation}
where $\Omega_d= \frac{2\pi^{d/2}}{\Gamma(d/2)}$ is the area of the unit sphere in $d$ dimension.
According to this definition, the function $g$ satisfies the boundary condition $g(k=0,\tau)=1$ and the
initial condition
\begin{equation*}
g_0(k)\equiv g(k,\tau=0)=\Omega_d\,\int_0^\infty dv\,v^{d-1}\,e^{-vk}\,f(v,\tau=0)
\end{equation*}
Applying the Laplace transform to Eq.~\eqref{final_v} yields
\begin{equation}
\label{transformed}
\frac{\partial g}{\partial\tau}=-d\,k\,g-k^2\,\frac{\partial g}{\partial k}
\end{equation}
The right hand side can be rewritten as $-k^{2-d}\,\frac{\partial}{\partial k} (k^d g)$ thereby 
suggesting to use the function $h=k^d g$ instead of $g$. One gets $h_\tau = -k^2 h_k$, 
or equivalently
\begin{equation}
\label{transformed2}
\frac{\partial h}{\partial \tau}=\frac{\partial h}{\partial \kappa}\,,\quad \kappa=k^{-1}.
\end{equation}
A general solution to the simple wave equation \eqref{transformed2} is $h(\kappa,\tau)=H(\kappa+\tau)$ where $H$ is determined by the initial condition: $h(\kappa,\tau=0)=H(\kappa)$. Returning to the original function $g$ we arrive at the general solution for the Laplace transform
\begin{equation}
\label{solution}
g(k,\tau)=(1+\tau k)^{-d}\,g_0\!\left(\frac{k}{1+k\tau}\right)
\end{equation}

As an example of the initial distribution with a compact support (that is, vanishing for sufficiently large velocities) consider the isotropic distribution with fixed initial speed $v_0$. In other words, let 
\begin{equation}
\label{in:uniform}
f(v,\tau=0)= \frac{\delta(v-v_0)}{\Omega_d \, v_0^{d-1}}
\end{equation}
In this case~\cite{large} the solution reads 
\begin{equation*}
g(k,\tau)=\frac{1}{(1+\tau k)^d}\, \exp\!\!\left[-\frac{v_0 k}{1+\tau k}\right]
\end{equation*}
Expanding the exponential and separately performing the inverse Laplace transform of each term we obtain
\begin{equation} 
f(v,\tau)=\frac{1}{\Omega_d \Gamma(d) \tau^d} \sum_{n=0}^{\infty} \frac{(-v_0/\tau)^n}{n!} \,
_1{F}_1\!\left[n+d;d;-\tfrac{v}{\tau}\right]
\label{exactsolution}
\end{equation}
where $_1{F}_1$ is the confluent hypergeometric function. 
The asymptotic behavior ($\tau \gg v_0 $) of \eqref{exactsolution}
is given by the first term ($n=0$) in the sum and is equal to 
\begin{equation}
f(v,\tau)=\frac{1}{\Omega_d \Gamma(d)} \, \frac{e^{-v/\tau}}{\tau^d} 
\label{asymptotic_solution}
\end{equation}
where we have used the identity $_1{F}_1[d;d;z]=e^z$.

As an example of an initial distribution with infinite support, consider an exponential distribution 
\begin{equation}
\label{in:exp}
f(v,\tau=0)=\frac{1}{\Omega_d \Gamma(d)} \, \frac{e^{-v/v_0}}{v_0^d} 
\end{equation}
In this case, the velocity distribution remains exponential throughout the evolution
\begin{equation}
\label{exp_in}
f(v,\tau)=\frac{1}{\Omega_d \Gamma(d)} \, \frac{e^{-v/(v_0+\tau)}}{(v_0+\tau)^d} 
\end{equation}
The asymptotic ($\tau \gg v_0$) behavior of the solution \eqref{exp_in}
is again given by \eqref{asymptotic_solution}. 

These two examples illustrate the general behavior which can be deduced from the general solution \eqref{solution}: If the initial velocity distribution decays exponentially or faster, the asymptotic behavior of the velocity distribution is universal (that is, independent on the initial velocity distribution) and given by \eqref{asymptotic_solution}. If the initial velocity distribution decays slower than exponentially in the $v\to\infty$ limit, the long time asymptotic behavior is given by Eq.~\eqref{asymptotic_solution} apart from the tail region. For instance, if $f(v,\tau=0)\sim v^{-\nu}$ as $v\to\infty$, the asymptotic velocity distribution is given by \eqref{asymptotic_solution} when $0\leq v\ll (\nu-d)\tau \ln\tau$, while for $v\gg (\nu-d)\tau \ln\tau$ the initial distribution dominates: $f(v,\tau)\sim v^{-\nu}$.

Essentially the same qualitative behavior is valid in the general case of the potential particle-atom interaction \eqref{Ur}. The governing kinetic equation \eqref{kinetic} describing the long time behavior is substantially more difficult than Eq.~\eqref{final_v} corresponding to the hard-sphere interaction, e.g. applying the Laplace transform to Eq.~\eqref{kinetic} does not lead to a closed equation for $g(k,\tau)$. Therefore it is much harder to prove rigorously that the asymptotic is given by \eqref{scaling-3d-form}--\eqref{scaling-3d-lambda}. A non-rigorous, but physically convincing, argument relies on the existence of a one-parameter family of {\em exact} solutions generalizing the scaling solution \eqref{scaling-3d-form}--\eqref{scaling-3d-lambda}. Indeed, let us start with an initial velocity distribution~\cite{large}  
\begin{equation}
\label{exp_init}
f(v,\tau=0)= \frac{C}{v_0^d}\,\exp\!\left\{ -\Lambda^2\,\left(\frac{v}{v_0}\right)^{1/\Lambda}\right\}
\end{equation}
where $v_0$ is a parameter and the constants $C$ and $\Lambda$ are the same as in Eqs.~\eqref{scaling-3d-form}--\eqref{scaling-3d-lambda}. A solution of Eq.~\eqref{kinetic} subject to the initial condition \eqref{exp_init} reads~\cite{l_exact}
\begin{equation}
\label{lambda_exact}
f=C\left(\tau+v_0^{1/\Lambda}\right)^{-\Lambda d}\,
\exp\!\left\{ -\Lambda^2\,\,\frac{v^{1/\Lambda}}{\tau+v_0^{1/\Lambda}}  \right\}
\end{equation}
Obviously, the velocity distribution \eqref{lambda_exact} approaches the scaling form \eqref{scaling-3d-form}--\eqref{scaling-3d-lambda} in the long time limit. This strongly suggests that for an arbitrary initial velocity distribution that decays as $\exp\!\left\{ -\text{const.}\times v^{1/\Lambda}\right\}$ or faster, the asymptotic behavior is given by \eqref{scaling-3d-form}--\eqref{scaling-3d-lambda}. For the initial velocity distribution decaying slower than the above stretched exponential, the asymptotic velocity distribution is still given by Eqs.~\eqref{scaling-3d-form}--\eqref{scaling-3d-lambda} in the major range and only the tail region is dominated by the initial velocity distribution.


\begin{thebibliography}{99}
\bibitem{even}
    The particle velocity distribution satisfies the condition $f(v,t)=f(-v,t)$ at all 
    times if the initial velocity distribution is even, e.g. $f(v,0)=\delta(v)$.  
    In the general case, the velocity distribution quickly becomes even.
\bibitem{calogero} 
    Apart from the hard-sphere gas, $\lambda=\infty$, the integration measure 
    is explicitly known for the Coulomb gas $\lambda=1$ (the Rutherford formula) 
    and the Calogero gas $\lambda=2$; in these three cases, one can compute
    the factor $A$.
\bibitem{MM_comment}
    The $r^{-2(d-1)}$ interaction potential characterizes Maxwell molecules 
    (which are actually atoms, the terminology reflects the tradition). 
    In our case we only require that the potential diverges as $r^{-2(d-1)}$ 
    in the small separation limit, and the potential actually describes 
    the particle-atom interaction; the atom-atom interaction is irrelevant. 
    In simulations, however, we usually assume that the particle-atom and 
    atom-atom interactions are equal.
\bibitem{sym}
    In the spatially homogeneous case the reflection symmetry is strictly 
    obeyed \cite{even} if it holds for the initial condition. In the 
    spatially inhomogeneous case it is not so, yet the asymmetry is weak, 
    e.g. the average velocity $\int_{-\infty}^\infty dv\,vf(x,v,t)$ remains bounded,
    while the average speed $\int_{-\infty}^\infty dv\,|v|f(x,v,t)$ grows as $\tau$. 
    Physically, in the large time limit when the typical velocity is large, the particle
    undergoes a great number of collisions so its velocity can be $\pm v$ with 
    almost the same probabilities.
\bibitem{diff_general}
    This expression reproduces correctly the known values of the diffusion coefficients for the 
    Lorentz gas in $d=1,2,3$ and we believe it holds in any $d$.
    In one dimension we recover $D=\frac{v}{2\rho}$ \cite{book}, while in three dimensions
    Eq.~\eqref{D_general} reduces to 
    $D=v/(3\pi a^2\rho)$, see \cite{fluid,LG-math,book}. The three-dimensional 
    formula is well-known. (Usually it is written in the form $D=v\ell/3$, with $\ell$
    being the mean-free path.) The expression (see e.g. Ref.~\cite{LG-math}) for the 
    diffusion coefficient in two dimensions, $D=3v/(16 a\rho)$, is less known since
    the hard-sphere scattering is non-isotropic in two dimensions, so in that 
    situation the Lorentz gas is harder to analyze than in three dimensions.
\bibitem{integer} 
    Equation \eqref{R4:d} for $\langle R^4\rangle$ shows that this moment is integer 
    when $d=1, 4, 7, 10, 13, \ldots$ (generally if $d=1+3p$ with an arbitrary 
    non-negative integer $p$) and non-integer in all other dimensions.  
    Equations \eqref{R4:d}--\eqref{R6:d} show that the 
    moments $\langle R^4\rangle$ and $\langle R^6\rangle$ are both integer when 
    $d=1, 13, 16, 28, 31, \ldots$ (generally when $d=1+15q$ or $d=13+15q$ with 
    an arbitrary non-negative integer $q$). Thus in some special dimensions a few
    first moments $\langle R^4\rangle$,  $\langle R^6\rangle$, etc. can be integer.
    It appears that in those dimensions only a few first moments are integer; the
    only exception is $d=1$ where all even moments are integer.
\bibitem{large} 
   We must assume that $v_0\gg \sqrt{T}$ in the initial distributions \eqref{in:uniform}
   and \eqref{in:exp} to ensure the validity of equation \eqref{final_v} during the entire
   time range $\tau>0$. The same inequality is required for the applicability of 
   Eq.~\eqref{kinetic} when the initial condition is given by Eq.~\eqref{exp_init}.
\bibitem{l_exact}
   Exact solutions \eqref{lambda_exact} easily follow from the scaling solution 
   \eqref{scaling-3d-form}--\eqref{scaling-3d-lambda} after noting that the 
   governing kinetic equation \eqref{kinetic} possesses the time-translational invariance.
\end{thebibliography}
\end{document}